\documentclass[%
reprint,
superscriptaddress,
 amsmath,amssymb,
 aps, physrev,
]{revtex4-2}

\usepackage{color}
\usepackage{xcolor}
\usepackage{tabularx}
\usepackage{array}

\newcolumntype{Y}{>{\centering\arraybackslash}X}
\usepackage{graphicx}
\usepackage{dcolumn}
\usepackage{siunitx}
\usepackage{bm}
\usepackage[version=4]{mhchem}
\usepackage{hyperref}
\hypersetup{colorlinks=true,linkcolor=blue,citecolor=blue,urlcolor=blue}
\usepackage{cleveref}
\usepackage[colorinlistoftodos]{todonotes}

\usepackage{soul} 

\begin{document}
\title{
Edelstein Effect in Ferromagnetic Monolayers \ce{Fe3GaTe2} and \ce{Fe3GeTe2}
}

\author{Tongrui Li}
\affiliation{Hefei National Research Center for Physical Sciences at Microscale and School of Emerging Technology, University of Science and Technology of China, Hefei, 230026, China}
\affiliation{National Synchrotron Radiation Laboratory, University of Science and Technology of China, Hefei, Anhui 230029, China}

\author{Lidong Zhang} 
\affiliation{National Synchrotron Radiation Laboratory, University of Science and Technology of China, Hefei, Anhui 230029, China}

\author{Weiyi Pan}
\email{weiyi.pan@physik.uni-regensburg.de}
\affiliation{Institute for Theoretical Physics, University of Regensburg, 93040 Regensburg, Germany}

\begin{abstract}
The Edelstein effect enables intrinsic current-induced spin polarization in noncentrosymmetric van der Waals magnets. Here, we systematically compare the Edelstein responses of monolayer \ce{Fe3GaTe2} and \ce{Fe3GeTe2} using first-principles Wannier-based linear-response calculations. Although both materials possess the same noncentrosymmetric $D_{3h}$ crystal symmetry and obey identical symmetry selection rules, their time-reversal-even and time-reversal-odd spin accumulations differ markedly in magnitude, sign, and energy dependence. Pronounced responses occur mainly on the hole-doped side near $-0.4$~eV in \ce{Fe3GaTe2}, but on the electron-doped side near $+0.4$~eV in \ce{Fe3GeTe2}. At the intrinsic Fermi level, atom-resolved decomposition shows that Ga-to-Ge substitution reverses the even contribution of the inequivalent Fe3 site for both in-plane magnetization directions and enhances its negative odd contribution, thereby reorganizing the cancellation among atomic sites. Momentum-resolved calculations associate the hole-side response of \ce{Fe3GaTe2} with a few large $\Gamma$-centered Fermi contours, whereas the electron-side response of \ce{Fe3GeTe2} involves a dense manifold of Fe-$3d$ bands that supports a strong interband response. The even response is determined by incomplete cancellation between opposite-sign intraband Fermi-surface contributions, whereas the odd response originates from interband coherence between closely spaced electronic states. Continuous magnetization rotation reveals symmetry-constrained angular dependences with appreciable higher-order harmonics, particularly in the $xz$ plane, while biaxial strain produces substantial amplitude variations and sign reversals. These results establish how crystal symmetry and material-specific electronic structure, together with magnetization orientation and lattice deformation, control the intrinsic Edelstein effect in the \ce{Fe3$X$Te2} ($X=\mathrm{Ga},\mathrm{Ge}$) family.
\end{abstract}

\maketitle
\clearpage
\section{Introduction}
The Edelstein effect converts charge transport into nonequilibrium spin polarization. In systems lacking spatial inversion symmetry, an applied electric field can produce a finite spin accumulation through this mechanism, which is also known as the inverse spin-galvanic effect~\cite{Edelstein1990,Johansson2024EdelsteinReview,xly2mhk9}. In magnetic materials, the exchange interaction couples the induced spin density to the magnetization, providing a microscopic source of spin--orbit torque (SOT) and enabling electrical control of magnetic order \cite{Manchon2019RMP,Fert2024ElectricalControl}. Most SOT devices demonstrated experimentally consist of a ferromagnet coupled to a heavy metal, topological material, or another layer with strong spin--orbit coupling \cite{Miron2011SOT,Liu2012TaSOT,Manchon2019RMP,Amin2020InterfacialSOT}. The torque in these heterostructures may arise from spin Hall transport, an interfacial Rashba--Edelstein response, or both~\cite{Manchon2019RMP,Amin2020InterfacialSOT}. Despite their effectiveness, the magnitude and character of these responses often depend sensitively on atomic intermixing, disorder, spin-memory loss, and interfacial hybridization~\cite{Amin2020InterfacialSOT,BaezFlores2022Intermixing,Gupta2020SpinMemoryLoss}. These interface-related effects motivate the study of intrinsically noncentrosymmetric magnets in which magnetism, spin--orbit coupling, and charge-to-spin conversion occur within a single crystalline material. Intrinsic current-induced torques have been observed in noncentrosymmetric ferromagnets without an additional spin-source layer \cite{Kurebayashi2014BerrySOT}. Such materials allow a more direct examination of the relationship among electronic structure, nonequilibrium spin accumulation, and magnetization dynamics.

Two-dimensional (2D) van der Waals (vdW) magnets offer a useful setting for investigating intrinsic current-induced spin phenomena \cite{Gong2017VDW,Huang2017CrI3,Tang2021VdWSOT}. Among metallic vdW ferromagnets, \ce{Fe3GeTe2} and \ce{Fe3GaTe2} are of particular interest because they combine strong perpendicular magnetic anisotropy (PMA) with efficient current-driven magnetic responses. Ferromagnetic order in \ce{Fe3GeTe2} persists down to the atomically thin limit \cite{Tan2018FGT,Fei2018FGT}, and its Curie temperature can be substantially enhanced by electrostatic gating \cite{Deng2018FGT}. In \ce{Fe3GaTe2}, the Curie temperature reaches approximately 350--380~K, accompanied by strong room-temperature PMA \cite{Zhang2022FGaT,Ruiz2024FGaT}. Hard ferromagnetism persists in monolayer \ce{Fe3GaTe2} \cite{Wang2024FGaTMonolayer}, and epitaxial one-unit-cell films retain perpendicular ferromagnetism above room temperature \cite{Wu2024FGaT}. Current-induced magnetization control has been demonstrated in Pt-based devices containing either material \cite{Alghamdi2019FGTSOT,Wang2019FGTSwitch,Li2023FGaTSOT,Kajale2024FGaTSwitch}, in all-vdW topological-insulator/\ce{Fe3GeTe2} heterostructures \cite{Wang2023FGTTI,Guillet2024FGTTI,Lohmann2024FGTTI}, and in single-material or homojunction devices \cite{Zhang2021FGTSingle,Lin2022FGTHomojunction,Yan2024FGaTSingle,Deng2024FGaTSingle}. Fully vdW heterostructures with low-symmetry \ce{WTe2} or \ce{TaIrTe4} spin-source layers have also shown efficient SOT switching and, in several cases, deterministic field-free reversal of \ce{Fe3GeTe2} and \ce{Fe3GaTe2} \cite{Shin2022WTe2FGT,Kao2022WTe2FGT,Li2024WTe2FGT,Kajale2024WTe2FGaT,Pandey2025TaIrTe4FGaT}. Related work has quantified intrinsic bulk SOTs in individual \ce{Fe3GeTe2} flakes \cite{Martin2023FGTBulkSOT}, demonstrated room-temperature self-switching in polar Co-substituted \ce{Fe_{2.5}Co_{2.5}GeTe2} \cite{Zhang2024PolarSelfSwitch}, realized orbital-torque-driven switching of \ce{Fe3GaTe2} \cite{Zhang2025FGaTOrbital}, and reported SOT-controlled nonvolatile switching in an all-vdW \ce{Fe3GaTe2}/\ce{Fe3GeTe2} bilayer \cite{Kang2026FGaTFGT}.


At the monolayer limit, \ce{Fe3GeTe2} and \ce{Fe3GaTe2} possess the same noncentrosymmetric $D_{3h}$ point-group symmetry. First-principles calculations for monolayer \ce{Fe3GeTe2} have shown that this symmetry allows intrinsic self-torques with substantial higher-order angular components and that these components can enable deterministic field-free switching of perpendicular magnetization~\cite{Xue2023FGT}. More recently, a direct comparison of monolayer \ce{Fe3GeTe2} and \ce{Fe3GaTe2} revealed markedly different atom- and momentum-resolved self-torques, despite their common symmetry and predominantly Fe-$3d$ electronic character. These differences were attributed to variations in band filling and Fermi-surface electronic structure~\cite{Brizolla2026FGTFGaTSOT}. Current-induced spin accumulation and SOT, however, are closely related but distinct quantities. Because the local torque is additionally weighted by the spatially varying exchange field, atom-resolved Edelstein responses do not necessarily follow atom-resolved torkances quantitatively \cite{Pan2026FGTIn2Se3}. This distinction raises a closely related microscopic question: how do the intrinsic Edelstein spin accumulations of \ce{Fe3GeTe2} and \ce{Fe3GaTe2} differ despite their identical point-group symmetry, and how are these differences encoded in their electronic structures?

In this work, we compare the  Edelstein responses of pristine monolayer \ce{Fe3GeTe2} and \ce{Fe3GaTe2} using density-functional-theory-based Wannier linear-response calculations. We first determine their electronic structures and identify the symmetry-allowed time-reversal-even and time-reversal-odd components for representative magnetization directions. We then compare the magnitudes, signs, and energy dependences of these components and use atom- and momentum-resolved calculations to trace their microscopic origins. Continuous rotation of the magnetization in the $xy$ and $xz$ planes reveals symmetry-constrained angular dependences and appreciable higher-order harmonics beyond the lowest-order forms. We also examine the modification of the Edelstein response by biaxial strain. Although \ce{Fe3GeTe2} and \ce{Fe3GaTe2} obey the same symmetry selection rules, their responses differ substantially in magnitude, sign, and characteristic energy scale because of their distinct low-energy electronic structures. These results clarify the relationship among crystal symmetry, electronic structure, and magnetization-dependent spin accumulation in the \ce{Fe3$X$Te2} ($X=\mathrm{Ga,Ge}$) family and provide a microscopic basis for controlling the spin-accumulation channel underlying current-induced torques in two-dimensional spintronic materials.

\section{Computational Methods}

The electronic structures of \ce{Fe3GaTe2} and \ce{Fe3GeTe2} were calculated using density functional theory (DFT) within the projector augmented-wave (PAW) method~\cite{prb5017}, as implemented in the Vienna \emph{ab initio} Simulation Package (VASP)~\cite{prb511169}. The exchange--correlation functional was treated within the local-density approximation (LDA)~\cite{PerdewZunger1981}. To suppress spurious interactions between periodically repeated layers along the out-of-plane direction, the lattice parameter along the $c$ axis was set to 30~\AA. Structural relaxations were performed in the collinear ferromagnetic state without spin--orbit coupling (SOC), using a plane-wave cutoff energy of 500~eV and a $20\times20\times1$ $\mathbf{k}$-point mesh. The atomic positions were relaxed until the residual Hellmann--Feynman forces on each atom were smaller than $10^{-3}$~eV/\AA. The optimized in-plane lattice constants were $a=b=3.948$~\AA\ for \ce{Fe3GaTe2} and $a=b=3.915$~\AA\ for \ce{Fe3GeTe2}.

After structural optimization, fully self-consistent calculations including SOC were carried out for different magnetization orientations. A denser $24\times24\times1$ $\mathbf{k}$-point mesh was employed for these calculations. Both in-plane and out-of-plane rotations of the magnetization were considered, and an independent self-consistent SOC calculation was performed for each magnetization direction.

To construct an accurate low-energy tight-binding representation, the Bloch states obtained from the SOC calculations were projected onto atomic-like orbitals using the \textsc{Wannier90} package~\cite{wannier90}. The Fe-$3d$, Ga/Ge-$p$, and Te-$p$ orbitals were adopted as the projection basis. Following the procedure used in previous first-principles studies of the Edelstein effect~\cite{Pan2026PtBi2,Pan2026VTe2}, no iterative maximal-localization procedure was applied, so that the orbital characters and symmetry properties of the original Bloch states were retained as closely as possible. The resulting Wannier Hamiltonians were further symmetrized using the \textsc{WannSymm} code~\cite{wannsymm}, which enforces the magnetic symmetry constraints on the Wannier tight-binding Hamiltonian.


The current-induced spin accumulation was calculated from the Wannier Hamiltonian using the \textsc{Linres} linear-response implementation~\cite{linres}. Within linear response, the induced spin accumulation $\delta \mathbf{S}$ is related to an external electric field $\mathbf{E}$ through the Edelstein susceptibility tensor $\boldsymbol{\chi}$,
\begin{equation}
    \delta S_i = \sum_j \chi_{ij} E_j .
\end{equation}
For a magnetic system, the Edelstein susceptibility can be decomposed into time-reversal-even and time-reversal-odd contributions,
\begin{equation}
    \chi_{ij}
    =
    \chi_{ij}^{\mathrm{even}}
    +
    \chi_{ij}^{\mathrm{odd}} ,
\end{equation}
which were evaluated within the Kubo linear-response formalism~\cite{Freimuth2014,Pan2026PtBi2,Pan2026VTe2}. The time-reversal-even contribution is given by
\begin{equation}
\chi_{ij}^{\mathrm{even}}
=
-\frac{e\hbar}{\pi N}
\sum_{\mathbf{k},m,n}
\frac{
\Gamma^2
\operatorname{Re}
\left[
\langle n\mathbf{k}|\hat{S}_i|m\mathbf{k}\rangle
\langle m\mathbf{k}|\hat{v}_j|n\mathbf{k}\rangle
\right]
}{
\left[(E_{\mathrm{F}}-\varepsilon_{n\mathbf{k}})^2+\Gamma^2\right]
\left[(E_{\mathrm{F}}-\varepsilon_{m\mathbf{k}})^2+\Gamma^2\right]
},
\label{eq:chi_even}
\end{equation}
whereas the time-reversal-odd contribution reads
\begin{equation}
\begin{split}
\chi_{ij}^{\mathrm{odd}}
=&
\frac{2e\hbar}{N}
\sum_{\mathbf{k}}
\sum_{n\in\mathrm{occ.}}
\sum_{\substack{m\in\mathrm{unocc.}\\m\neq n}}
\operatorname{Im}
\left[
\langle n\mathbf{k}|\hat{S}_i|m\mathbf{k}\rangle
\langle m\mathbf{k}|\hat{v}_j|n\mathbf{k}\rangle
\right]
\\
&\times
\frac{
\Gamma^2-
(\varepsilon_{n\mathbf{k}}-\varepsilon_{m\mathbf{k}})^2
}{
\left[
(\varepsilon_{n\mathbf{k}}-\varepsilon_{m\mathbf{k}})^2
+\Gamma^2
\right]^2
}.
\label{eq:chi_odd}
\end{split}
\end{equation}
Here, $e$ is the elementary charge, $\hbar$ is the reduced Planck constant, $N$ is the total number of $\mathbf{k}$ points, $E_{\mathrm{F}}$ is the Fermi energy, $\varepsilon_{n\mathbf{k}}$ denotes the band energy, $\hat{S}_i$ and $\hat{v}_j$ are the spin and velocity operators, respectively, and $\Gamma$ represents the phenomenological broadening. The latter is related to the relaxation time through $\tau=\hbar/(2\Gamma)$. We adopted $\Gamma=10$~meV, following Refs.~\cite{Pan2026PtBi2,Pan2026VTe2}, which corresponds to a relaxation time of approximately $\tau=33$~fs. The dependence of the calculated Edelstein response on the broadening parameter $\Gamma$ is examined in detail in Appendix~B. A dense $600\times600\times1$ $\mathbf{k}$-point mesh was used for the linear-response calculations to ensure convergence of the Edelstein coefficients. In the present calculations, the electric field is applied along the $+x$ direction, and therefore only the first column of the Edelstein tensor is required, with $\delta S_i=\chi_{ix}E_x$.

Carrier-filling effects were modeled within the rigid-band approximation by varying the chemical potential relative to the intrinsic Fermi level while keeping the underlying electronic structure fixed. The biaxial strain is defined as \(\varepsilon=(a-a_0)/a_0\), where \(a_0\) is the equilibrium in-plane lattice constant. For each strain value between \(-2\%\) and \(+2\%\), the in-plane lattice constant was fixed to \(a=(1+\varepsilon)a_0\), while (the internal atomic coordinates were fully relaxed). The electronic structure and Wannier Hamiltonian were then recalculated for each strained configuration.

\begin{figure}[t]
\includegraphics[width=\columnwidth]{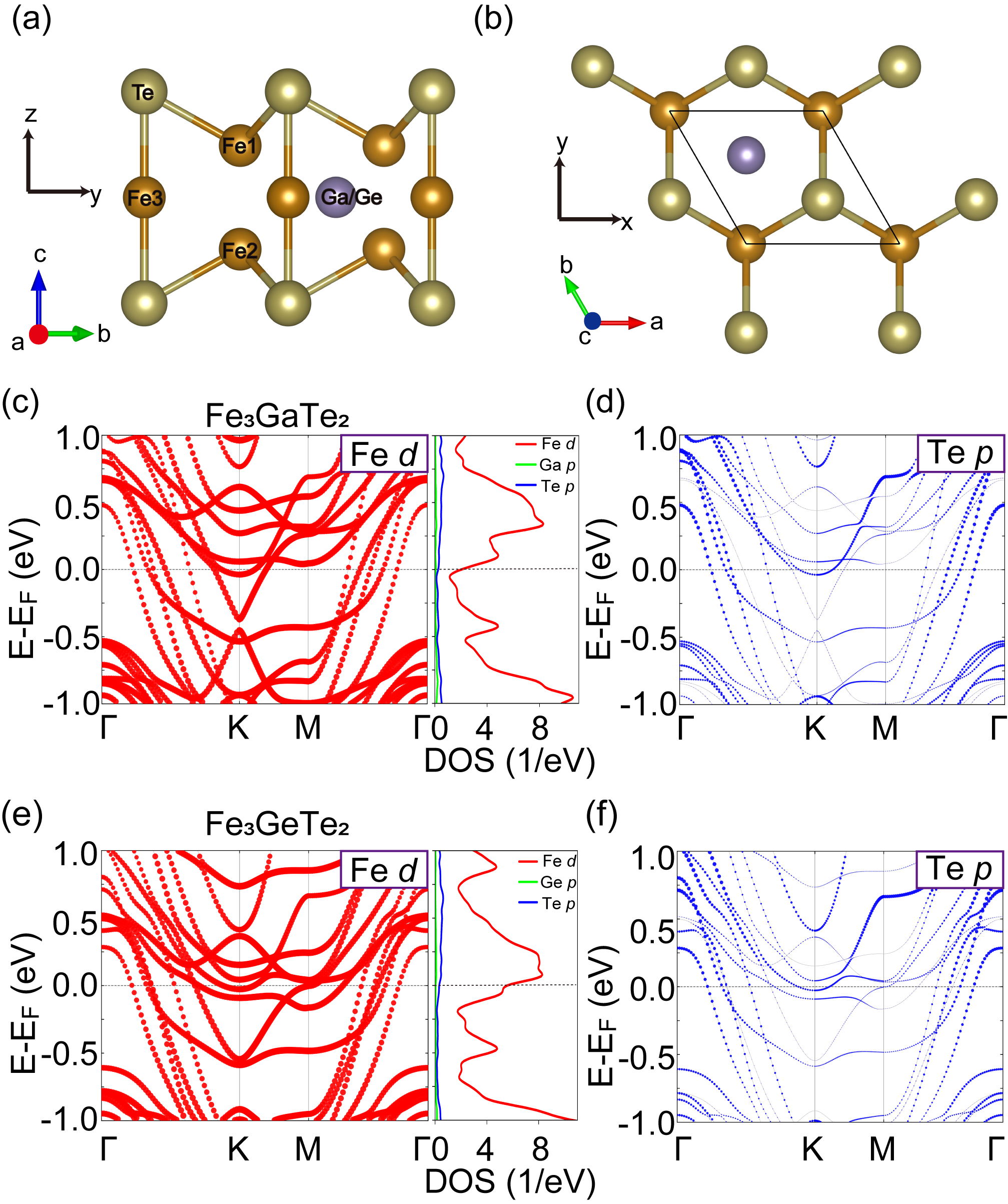}
\caption{ (a,b) Side and top views of the crystal structures of monolayer \ce{Fe3GaTe2} and \ce{Fe3GeTe2}. The inequivalent Fe sites are labeled as Fe1, Fe2, and Fe3; Fe1 and Fe2 are related by crystal symmetry, and the Te atoms form symmetry-equivalent pairs. (c,d) Orbital-projected band structures of monolayer \ce{Fe3GaTe2} with Fe-$d$ and Te-$p$ characters highlighted, together with the corresponding orbital-resolved DOS. (e,f) Orbital-projected band structures and DOS of monolayer \ce{Fe3GeTe2}. All calculations are performed with the magnetization oriented along the out-of-plane $c$ axis. }
\label{fig1}
\end{figure}

\section{Results and Discussion}
\subsection{Electronic structure and symmetry analysis}

The crystal structures of monolayer \ce{Fe3GaTe2} and \ce{Fe3GeTe2} are shown in Figs.~\ref{fig1}(a) and \ref{fig1}(b), respectively, from the side and top views. Both compounds crystallize in the hexagonal space group $P\bar{6}m2$ (No.~187), corresponding to the $D_{3h}$ point group. Each primitive cell contains three Fe atoms occupying two crystallographically distinct sites, with Fe1 and Fe2 being symmetry equivalent, together with one Ga/Ge atom and two symmetry-equivalent Te atoms. The side view highlights the layered arrangement of the \ce{Fe3$X$Te2} ($X=$ Ga, Ge) monolayer, whereas the top view displays the threefold rotational symmetry of the lattice. The magnetic anisotropy calculations show that both monolayers favor out-of-plane magnetization. Taking the magnetic anisotropy energy as the energy difference between the in-plane and out-of-plane configurations, the out-of-plane state is lower in energy by approximately $0.37$~meV per Fe atom in \ce{Fe3GaTe2} and $1.0$~meV per Fe atom in \ce{Fe3GeTe2}, indicating perpendicular magnetic anisotropy in both systems. Figures~\ref{fig1}(c)--\ref{fig1}(f) present the orbital-projected band structures and corresponding projected densities of states (DOS). In both \ce{Fe3GaTe2} and \ce{Fe3GeTe2}, the states near the Fermi level are dominated by Fe-$3d$ orbitals, with weaker Te-$p$ contributions arising from hybridization with the Fe-derived bands, consistent with previous calculations for monolayer \ce{Fe3GaTe2} \cite{Ruiz2024FGaT,5c00977} and \ce{Fe3GeTe2} \cite{Xue2023FGT}. Despite their common crystal symmetry and similar orbital character, the two compounds exhibit markedly different band dispersions and Fermi-surface structures near the Fermi level, reflecting the change in band filling induced by the Ga--Ge substitution \cite{Brizolla2026FGTFGaTSOT}. They therefore provide a useful platform for distinguishing symmetry-imposed features of the current-induced response from those governed by material-specific electronic structure.

Having established the basic electronic structures, we next examine the magnetization-dependent Edelstein response from the viewpoint of magnetic symmetry. Both monolayers possess the same noncentrosymmetric $D_{3h}$ crystal symmetry, but the magnetic symmetry is reduced once a specific magnetization direction is chosen. The symmetry-allowed components can then be identified from the transformation properties of $E_x$ and $S_i$. For a unitary spatial operation $R$, each tensor component $\chi_{ix}$ transforms according to the signs acquired by $S_i$ and $E_x$. For an antiunitary operation $RT$, the time-reversal-even response follows the same spatial transformation, whereas the time-reversal-odd response acquires an additional minus sign, because
\[
\chi_{ij}^{\mathrm{even}}(-\mathbf{m})
=
\chi_{ij}^{\mathrm{even}}(\mathbf{m}),
\qquad
\chi_{ij}^{\mathrm{odd}}(-\mathbf{m})
=
-\chi_{ij}^{\mathrm{odd}}(\mathbf{m}).
\]

\begin{table}[t]
\caption{\label{tab:symmetry_selection}
Symmetry constraints on the first column of the Edelstein tensor for an electric field along $+x$. The entries $(s_{xx},s_{yx},s_{zx})$ give the transformation signs of $(\chi_{xx},\chi_{yx},\chi_{zx})$.}
\centering
\renewcommand{\arraystretch}{1.15}
\setlength{\tabcolsep}{4pt}
\begin{tabular}{c c c c}
\hline\hline
$\mathbf{m}$ & Operation
& $\chi^{\mathrm{even}}_{ix}$
& $\chi^{\mathrm{odd}}_{ix}$ \\
\hline
$+x$ & $M_x$
& $(-,+,+)$ & $(-,+,+)$ \\
     & $C_{2y}T$
& $(+,-,+)$ & $(-,+,-)$ \\
     & $M_zT$
& $(-,-,+)$ & $(+,+,-)$ \\
\cline{2-4}
     & Allowed
& $\chi_{zx}^{\mathrm{even}}$
& $\chi_{yx}^{\mathrm{odd}}$ \\
\hline
$+y$ & $C_{2y}$
& $(+,-,+)$ & $(+,-,+)$ \\
     & $M_xT$
& $(-,+,+)$ & $(+,-,-)$ \\
     & $M_zT$
& $(-,-,+)$ & $(+,+,-)$ \\
\cline{2-4}
     & Allowed
& $\chi_{zx}^{\mathrm{even}}$
& $\chi_{xx}^{\mathrm{odd}}$ \\
\hline\hline
\end{tabular}
\end{table}

For $\mathbf{m}\parallel+x$, the magnetic point group is generated by $\{E,M_x,C_{2y}T,M_zT\}$. As an example, under $M_x$, $E_x\rightarrow-E_x$, while the axial spin transforms as $(S_x,S_y,S_z)\rightarrow(S_x,-S_y,-S_z)$. Consequently, $\chi_{xx}$ changes sign whereas $\chi_{yx}$ and $\chi_{zx}$ remain invariant, giving the first row of Table~\ref{tab:symmetry_selection}. The operation $C_{2y}T$ further distinguishes the even and odd responses: it eliminates $\chi_{yx}^{\mathrm{even}}$ but preserves $\chi_{zx}^{\mathrm{even}}$, whereas in the odd sector it preserves $\chi_{yx}^{\mathrm{odd}}$ but eliminates $\chi_{zx}^{\mathrm{odd}}$. Finally, $M_zT$ leaves the two surviving components invariant. Therefore, the symmetry allowed components are:
\begin{equation}
\mathbf{m}\parallel+x:
\qquad
\delta S_z^{\mathrm{even}}
=
\chi_{zx}^{\mathrm{even}}E_x,
\qquad
\delta S_y^{\mathrm{odd}}
=
\chi_{yx}^{\mathrm{odd}}E_x .
\label{eq:allowed_Mx}
\end{equation}

For $\mathbf{m}\parallel+y$, the magnetic point group is instead generated by $\{E,C_{2y},M_xT,M_zT\}$. The unitary $C_{2y}$ operation first excludes $\chi_{yx}$, while the two antiunitary operations impose different constraints on the even and odd sectors. As summarized in Table~\ref{tab:symmetry_selection}, only $\chi_{zx}^{\mathrm{even}}$ remains invariant in the even sector, whereas only $\chi_{xx}^{\mathrm{odd}}$ survives in the odd sector. Hence,
\begin{equation}
\mathbf{m}\parallel+y:
\qquad
\delta S_z^{\mathrm{even}}
=
\chi_{zx}^{\mathrm{even}}E_x,
\qquad
\delta S_x^{\mathrm{odd}}
=
\chi_{xx}^{\mathrm{odd}}E_x .
\label{eq:allowed_My}
\end{equation}
These selection rules are characteristic of the magnetic $D_{3h}$ symmetry and coincide with those obtained for monolayer $2H$-\ce{VTe2}, where the same time-reversal-even and time-reversal-odd spin components are allowed for $\mathbf{m}\parallel+x$ and $\mathbf{m}\parallel+y$ \cite{Pan2026VTe2}. They are also consistent with the lowest-order symmetry expansion of current-induced spin polarization in magnetic transition-metal dichalcogenides with $D_{3h}$ symmetry \cite{Li2020PNAS}.

\begin{figure}[htpb]
\includegraphics[width=8.5cm]{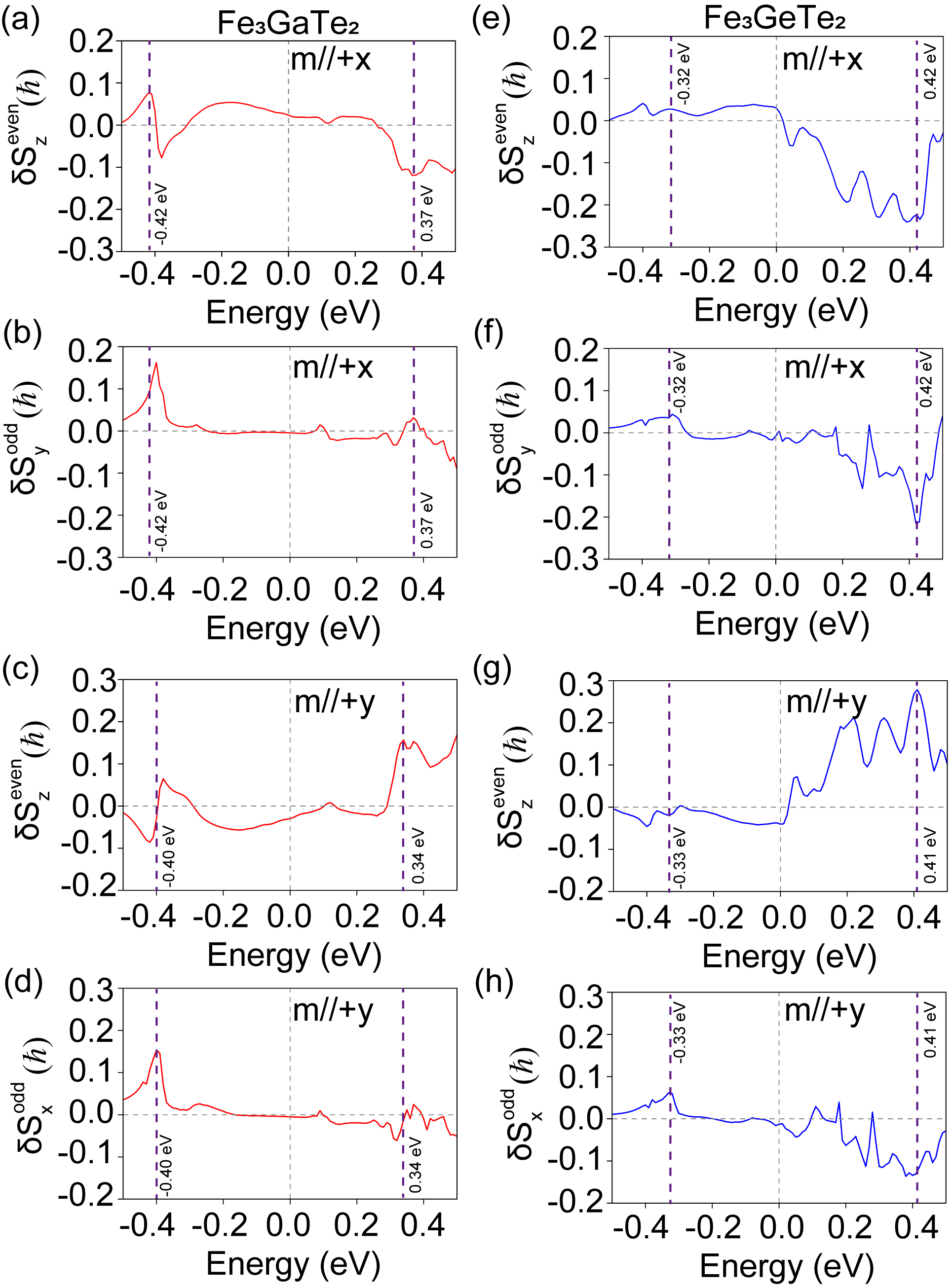}
\caption{ Calculated current-induced spin accumulations as functions of the Fermi energy for monolayer \ce{Fe3GaTe2} and \ce{Fe3GeTe2} under an applied electric field of 1~V/\AA. Panels (a)--(d) present the results for \ce{Fe3GaTe2}, while panels (e)--(h) show the corresponding results for \ce{Fe3GeTe2}. For magnetization oriented along the $+x$ direction, panels (a,e) and (b,f) display the time-reversal-even $\delta S_{z}^{\mathrm{even}}$ and time-reversal-odd $\delta S_{y}^{\mathrm{odd}}$ components, respectively. For magnetization along the $+y$ direction, panels (c,g) and (d,h) show the corresponding $\delta S_{z}^{\mathrm{even}}$ and $\delta S_{x}^{\mathrm{odd}}$ components. The purple dashed lines indicate selected characteristic energies where pronounced features appear in the Edelstein response, which are further analyzed by momentum-resolved spin accumulation calculations. }
\label{fig2}
\end{figure}

The magnetic symmetry also constrains the site-resolved Edelstein response. As illustrated in Fig.~\ref{fig1}(a), Fe1 and Fe2 form a pair of crystallographically equivalent sites related by the horizontal mirror $M_z$, and the two Te atoms are related in the same manner. For the in-plane magnetic configurations considered above, the corresponding magnetic symmetry is $M_zT$: its spatial part exchanges Fe1 and Fe2, as well as the upper and lower Te sites, while time reversal leaves their atomic positions unchanged. Importantly, all symmetry-allowed response components identified above, $\chi_{zx}^{\mathrm{even}}$, $\chi_{yx}^{\mathrm{odd}}$ for $\mathbf{m}\parallel+x$, and $\chi_{xx}^{\mathrm{odd}}$ for $\mathbf{m}\parallel+y$, are invariant under $M_zT$. The two equivalent Fe sites must therefore contribute equally to these allowed spin-accumulation components, and the same relation holds for the two equivalent Te sites. Accordingly, in the atom-resolved analysis below, Fe1 and Te1 may be used to represent the respective symmetry-equivalent pairs, with their multiplicities included when evaluating the total response. The above analysis concerns the two high-symmetry in-plane magnetization directions. When the magnetization is continuously rotated within the $xy$ plane, the additional symmetries specific to $\mathbf{m}\parallel+x$ or $+y$ are progressively reduced, and both in-plane time-reversal-odd components can in general become finite, while their angular dependences remain constrained by the underlying $D_{3h}$ crystal symmetry. The full magnetization-angle dependence, including the corresponding higher-order angular harmonics, is discussed later in connection with Fig.~~\ref{fig6} and derived in detail in Appendix~\ref{app:angular}.

\subsection{Energy-dependent Edelstein response}

With these symmetry constraints established, Fig.~\ref{fig2} presents
the calculated energy-dependent spin accumulations of the two
monolayers under a reference electric field of $1$~V/\AA\ applied
along $+x$. The response of \ce{Fe3GaTe2} varies strongly with both
energy and magnetization direction. For $\mathbf{m}\parallel +x$,
Fig.~\ref{fig2}(a) shows that $\delta S_z^{\mathrm{even}}$ reaches
approximately $0.10\,\hbar$ at $E-E_F\simeq-0.42$~eV. It then changes
sign on the electron side and decreases to approximately
$-0.12\,\hbar$ near $+0.37$~eV. As shown in
Fig.~\ref{fig2}(b), the corresponding time-reversal-odd component
$\delta S_y^{\mathrm{odd}}$ exhibits a sharper positive peak of about
$0.16\,\hbar$ near $-0.40$~eV. Upon rotating the magnetization to
$+y$, $\delta S_z^{\mathrm{even}}$ develops two extrema of opposite
sign in Fig.~\ref{fig2}(c): a negative extremum of approximately
$-0.10\,\hbar$ near $-0.40$~eV and a positive maximum of about
$0.16\,\hbar$ near $+0.34$~eV. Figure~\ref{fig2}(d) further shows that
$\delta S_x^{\mathrm{odd}}$ has a pronounced positive peak of
approximately $0.15\,\hbar$ on the hole side before becoming negative
on the electron side.

\begin{figure*}[htpb]
\includegraphics[width=17cm]{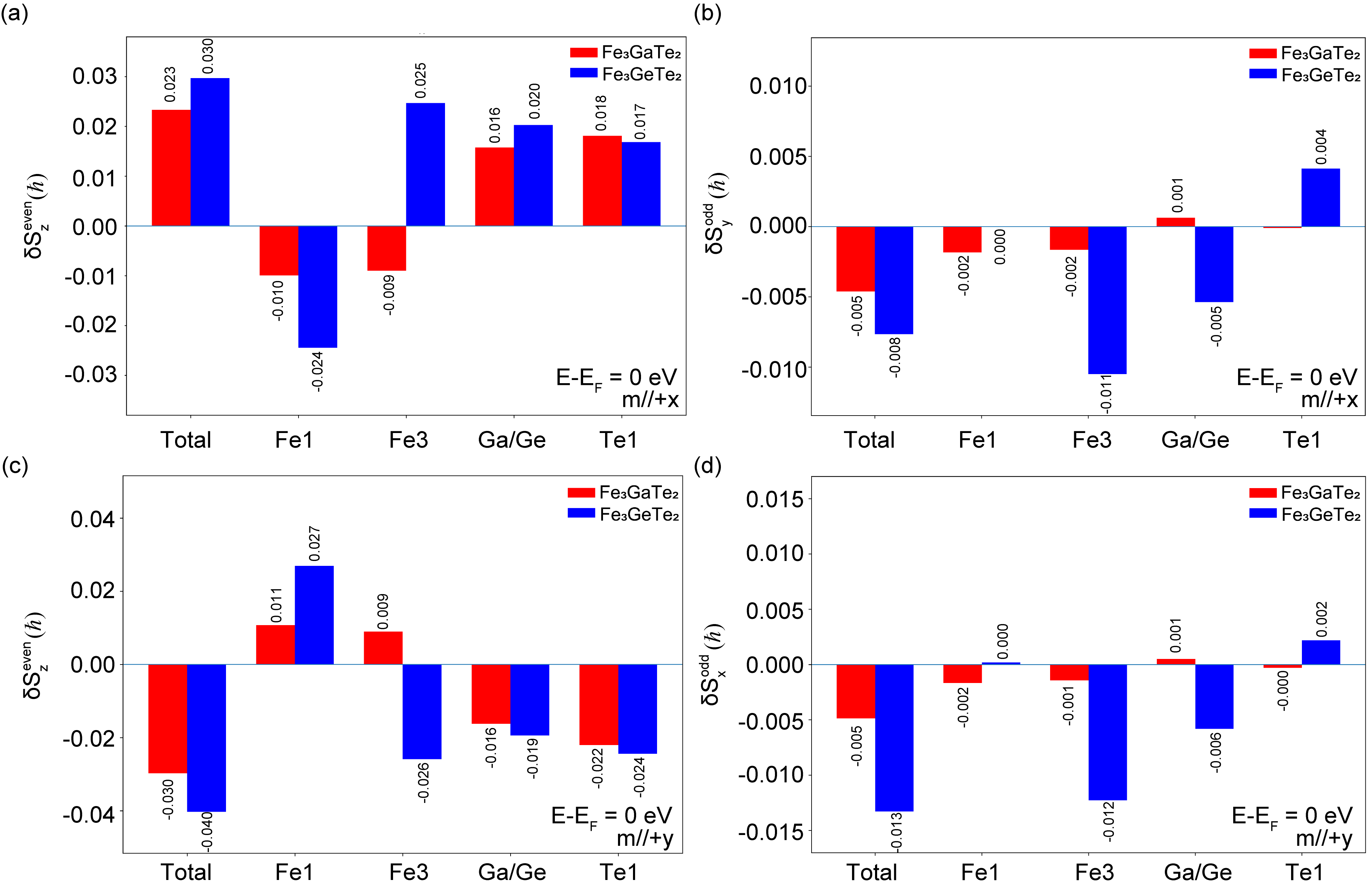}
\caption{ Comparison of the total and atom-resolved current-induced spin accumulations at the Fermi level ($E-E_{\mathrm F}=0$) for monolayer \ce{Fe3GaTe2} and \ce{Fe3GeTe2} under an applied electric field of $E_x=1$~V/\AA. (a,b) Time-reversal-even $\delta S_{z}^{\mathrm{even}}$ and time-reversal-odd $\delta S_{y}^{\mathrm{odd}}$, respectively, for magnetization along the $+x$ direction. (c,d) Time-reversal-even $\delta S_{z}^{\mathrm{even}}$ and time-reversal-odd $\delta S_{x}^{\mathrm{odd}}$, respectively, for magnetization along the $+y$ direction. The bars labeled Fe1, Fe3, Ga/Ge, and Te1 denote the contributions from the corresponding atomic sites. Fe1 and Te1 each represent one of two symmetry-equivalent Fe and Te sites, respectively; their multiplicities are included in the total spin accumulation. Red and blue bars correspond to \ce{Fe3GaTe2} and \ce{Fe3GeTe2}, respectively. }
\label{fig3}
\end{figure*}
The corresponding responses of \ce{Fe3GeTe2} exhibit a distinctly
different energy dependence. On the hole side around $-0.3$~eV, all
four components remain relatively modest, with characteristic
magnitudes below approximately $0.06\,\hbar$. Much stronger responses
emerge above the Fermi level. Figures~\ref{fig2}(e) and
\ref{fig2}(f) show that, for $\mathbf{m}\parallel +x$, both
$\delta S_z^{\mathrm{even}}$ and $\delta S_y^{\mathrm{odd}}$ become
large and negative around $E-E_F\simeq+0.4$~eV, reaching magnitudes of
approximately $0.20$--$0.25\,\hbar$. For
$\mathbf{m}\parallel +y$, Fig.~\ref{fig2}(g) shows that
$\delta S_z^{\mathrm{even}}$ instead develops a large positive peak
approaching $0.30\,\hbar$. Over the same energy range,
$\delta S_x^{\mathrm{odd}}$ reaches approximately
$-0.15\,\hbar$, as shown in Fig.~\ref{fig2}(h).

Several features can therefore be identified from Fig.~\ref{fig2}. First, the set of nonzero spin components is fixed by magnetic symmetry and is therefore the same in \ce{Fe3GeTe2} and \ce{Fe3GaTe2}. Their magnitudes, signs, and characteristic energy scales, however, differ markedly. In particular, \ce{Fe3GaTe2} exhibits its strongest responses mainly on the hole-doped side around $-0.4$~eV, whereas the largest responses in \ce{Fe3GeTe2} occur predominantly on the electron-doped side near $+0.4$~eV. This pronounced material dependence is consistent with recent torque calculations showing that the Ga--Ge substitution strongly modifies the band filling and Fermi-surface electronic structure of the \ce{Fe3X$$Te2} family~\cite{Brizolla2026FGTFGaTSOT}. More generally, current-induced responses in \ce{Fe3GeTe2} have been found to be highly sensitive to carrier filling, which can reorganize the relative contributions of different microscopic torque channels~\cite{Saunderson2022FGT}. A similarly strong Fermi-level dependence occurs for the Edelstein effect in monolayer \ce{PtBi2}, where competing Fermi-surface pockets contribute with opposite signs and changes in their relative weights can reverse the net response \cite{Pan2026PtBi2}. These comparisons support the general picture that the quantitative Edelstein response is governed by the detailed low-energy electronic structure rather than by symmetry alone. At the intrinsic Fermi level, the responses of both compounds are smaller but remain finite, providing a convenient reference for comparing their microscopic origins. We therefore next decompose the Fermi-level spin accumulation into atom-resolved contributions.

\begin{figure*}[htpb]
\includegraphics[width=18cm]{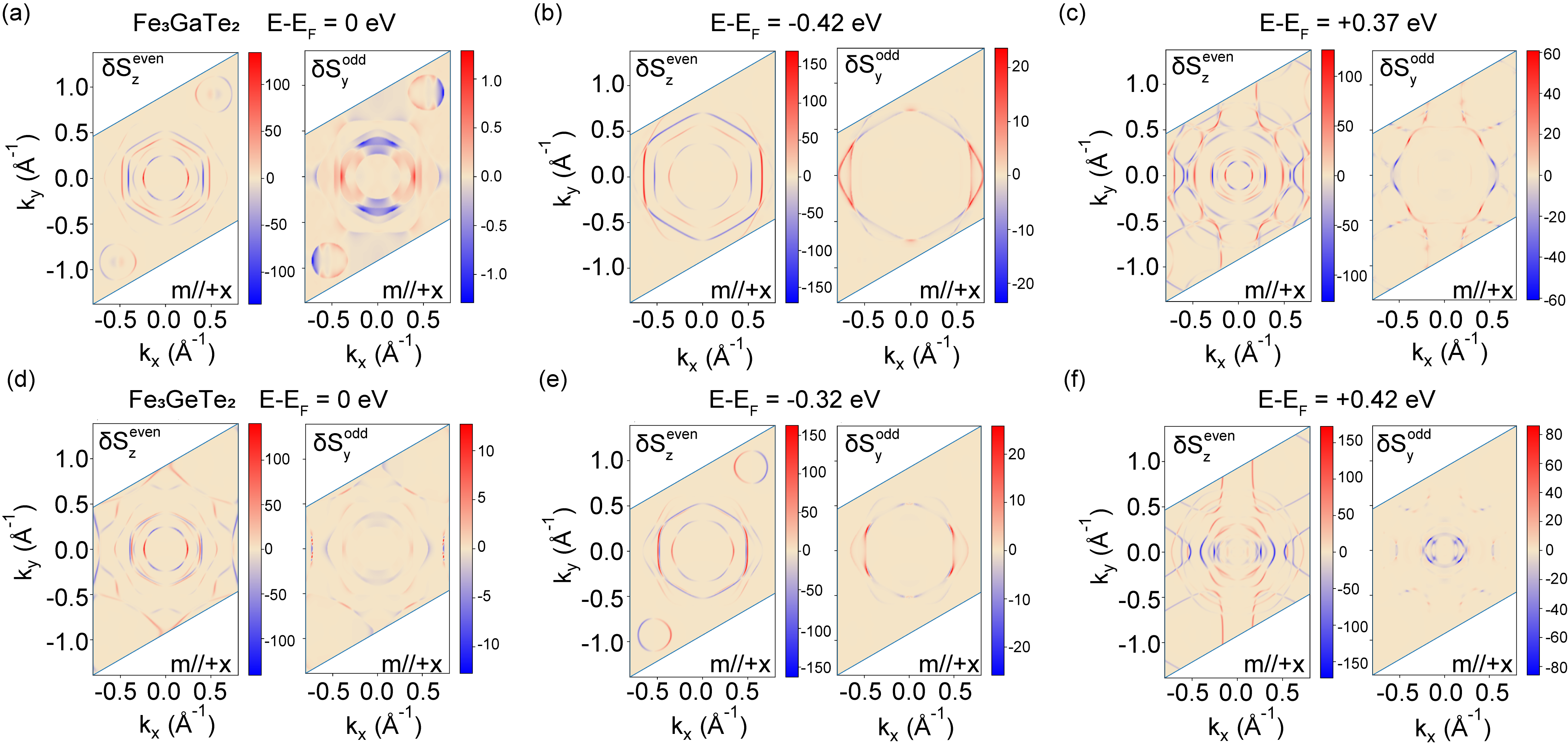}
\caption{Momentum-resolved current-induced spin accumulations in monolayer
\ce{Fe3GaTe2} and \ce{Fe3GeTe2} for magnetization along the $+x$ direction under a reference electric field of 1~V/\AA\ applied along $+x$. Panels (a)--(c) show the results for \ce{Fe3GaTe2} at $E-E_{\mathrm F}=0$, $-0.42$, and $+0.37$~eV, respectively, whereas panels (d)--(f) show those for \ce{Fe3GeTe2} at $E-E_{\mathrm F}=0$, $-0.32$, and $+0.42$~eV, respectively. For each energy, the left and right subpanels display the time-reversal-even $\delta S_{z}^{\mathrm{even}}(\mathbf{k})$ and time-reversal-odd $\delta S_{y}^{\mathrm{odd}}(\mathbf{k})$ components, respectively. The selected energies correspond to representative extrema of the energy-dependent spin accumulations shown in Fig.~\ref{fig2}. Red and blue indicate positive and negative contributions to the momentum-resolved spin accumulation, respectively. The momentum-resolved spin accumulation is given in units of \(\hbar\,\text{\AA}^3\).}
\label{fig4}
\end{figure*}

\subsection{Microscopic origin of the Edelstein response}

To gain further insight into the microscopic origin of the Edelstein response, Fig.~\ref{fig3} decomposes the Fermi-level spin accumulation into contributions from different atomic sites. Because Fe1 and Fe2 are symmetry equivalent and the two Te atoms form another equivalent pair, only one representative Fe1 and Te1 site is displayed. Their multiplicities are nevertheless included in the total response. Figures~\ref{fig3}(a) and \ref{fig3}(c) show that the two compounds have comparable total time-reversal-even responses but markedly different microscopic decompositions. For $\mathbf{m}\parallel+x$, $\delta S_z^{\mathrm{even}}$ is positive in both materials, whereas it becomes negative when the magnetization is rotated to $+y$. In \ce{Fe3GaTe2}, the net response results from substantial cancellation among the Fe, Ga, and Te contributions. In \ce{Fe3GeTe2}, this balance is strongly reorganized. The Fe3 contribution displays a particularly pronounced change in both magnitude and sign as the magnetization direction varies. These results show that replacing Ga with Ge modifies not only the response of the substituted site but also the current-induced spin polarization of the Fe-derived states.

Figures~\ref{fig3}(b) and \ref{fig3}(d) show that the time-reversal-odd responses are smaller at the Fermi level but retain a clear material dependence. In \ce{Fe3GeTe2}, the Fe3 site makes a sizable contribution to the odd response, accompanied by substantial Ge and Te contributions. The corresponding atomic contributions are considerably weaker in \ce{Fe3GaTe2}. As a result, the total odd response is larger in \ce{Fe3GeTe2}, particularly for $\mathbf{m}\parallel+y$. Taken together, the atom-resolved results show that the Edelstein response arises from competing contributions from several atomic sites rather than from a single dominant species. Replacing Ga with Ge substantially redistributes the response among the Fe sublattices, with Fe3 exhibiting the strongest material dependence. Such site-dependent and partially compensating contributions are consistent with recent atom-resolved calculations of self-torques in monolayer \ce{Fe3GeTe2} and \ce{Fe3GaTe2} \cite{Brizolla2026FGTFGaTSOT}. The connection between spin accumulation and torque should, nevertheless, be interpreted with care. Calculations for \ce{Fe3GeTe2}/\ce{In2Se3} have shown that the atom-resolved Edelstein spin accumulation can correlate qualitatively with the local torkance, but the two need not agree quantitatively because the torque is additionally weighted by the spatially nonuniform exchange field \cite{Pan2026FGTIn2Se3}. The present decomposition therefore provides information complementary to the local torkance by resolving the underlying nonequilibrium spin density, but it does not uniquely determine the torque.

\begin{figure*}[htpb]
\includegraphics[width=18cm]{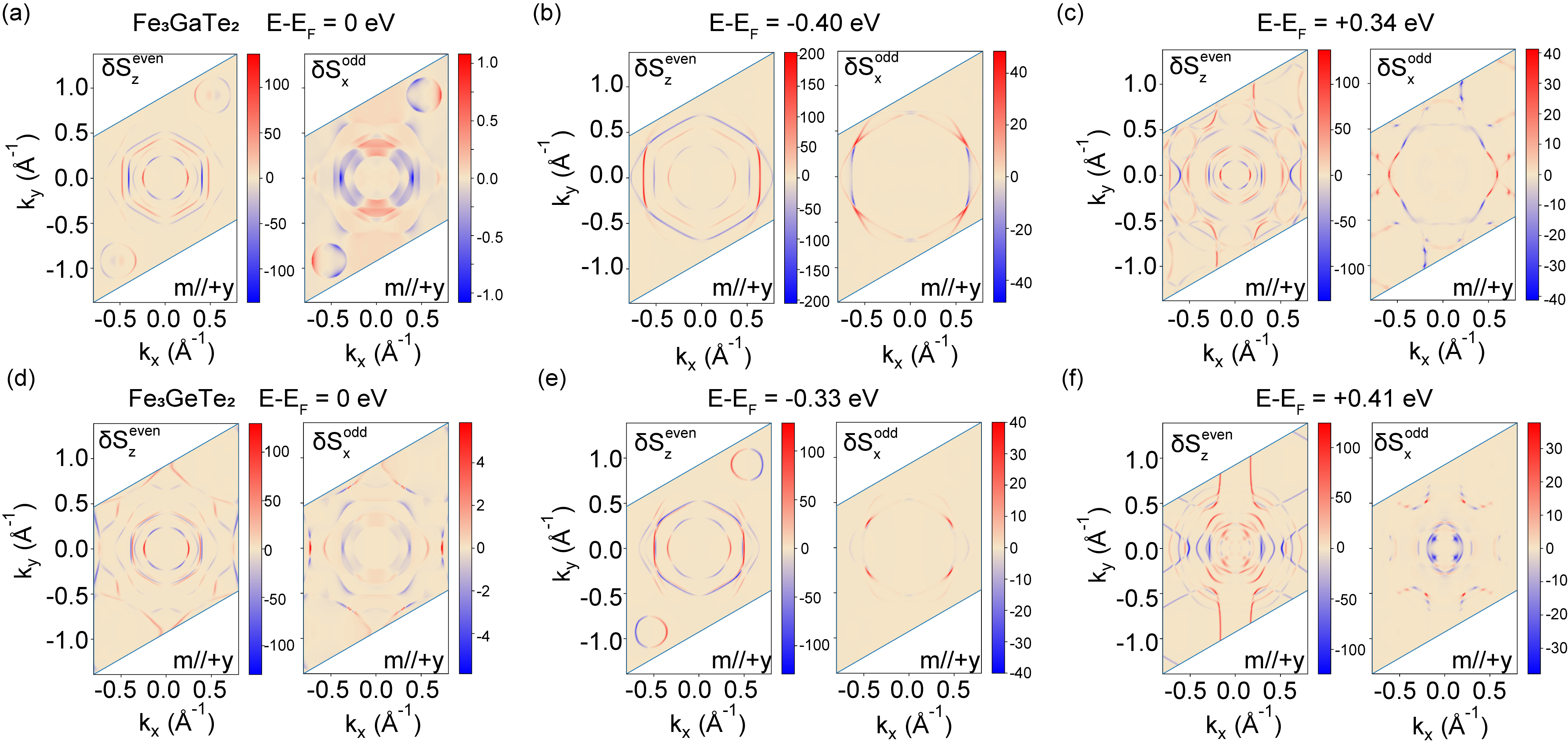}
\caption{Momentum-resolved current-induced spin accumulations in monolayer
\ce{Fe3GaTe2} and \ce{Fe3GeTe2} for magnetization along the $+y$ direction under a reference electric field of 1~V/\AA\ applied along $+x$. Panels (a)--(c) show the results for \ce{Fe3GaTe2} at $E-E_{\mathrm F}=0$, $-0.40$, and $+0.34$~eV, respectively, whereas panels (d)--(f) show those for \ce{Fe3GeTe2} at $E-E_{\mathrm F}=0$, $-0.33$, and $+0.41$~eV, respectively. For each energy, the left and right subpanels display the time-reversal-even $\delta S_{z}^{\mathrm{even}}(\mathbf{k})$ and time-reversal-odd $\delta S_{x}^{\mathrm{odd}}(\mathbf{k})$ components, respectively. The selected energies correspond to representative extrema of the energy-dependent spin accumulations shown in Fig.~\ref{fig2}. Red and blue denote positive and negative contributions, respectively. The momentum-resolved spin accumulation is given in units of $\hbar\,\mathrm{\AA}^{3}$.}
\label{fig5}
\end{figure*}

The atomic analysis does not identify the electronic states responsible for the response in momentum space. We therefore calculate the momentum-resolved spin accumulation at the Fermi level and at representative energies corresponding to the characteristic extrema in Fig.~\ref{fig2}. The results for $\mathbf{m}\parallel+x$ and $\mathbf{m}\parallel+y$ are shown in Figs.~\ref{fig4} and \ref{fig5}, respectively, while the corresponding band structures and selected energies are summarized in Fig.~\ref{fig8}. In the small-broadening limit, the time-reversal-even response is dominated by the intraband contribution and is therefore concentrated near the Fermi surface, whereas the time-reversal-odd response is interband in origin and is enhanced where spin--velocity-coupled bands approach one another in energy \cite{Pan2026VTe2,Pan2026PtBi2}.

For $\mathbf{m}\parallel+x$, Figs.~\ref{fig4}(a) and \ref{fig4}(d) show that the Fermi-level distributions of \ce{Fe3GaTe2} and \ce{Fe3GeTe2} contain pronounced positive and negative contributions along several $\Gamma$-centered momentum-space contours. Their substantial cancellation results in a relatively modest net response. At the representative hole-side energies, $E-E_F=-0.42$~eV for \ce{Fe3GaTe2} and $-0.32$~eV for \ce{Fe3GeTe2}, the time-reversal-even response is concentrated on a smaller number of large Fermi contours, as shown in Figs.~\ref{fig4}(b) and \ref{fig4}(e). This simpler distribution is consistent with the smaller number of band crossings at these energies in Figs.~\ref{fig8}(a) and \ref{fig8}(c). At the corresponding positive energies of $+0.37$~eV and $+0.42$~eV, multiple Fe-$d$-derived bands contribute simultaneously. The resulting momentum-space distributions in Figs.~\ref{fig4}(c) and \ref{fig4}(f) therefore exhibit more complex sign-alternating patterns. This behavior is particularly pronounced in \ce{Fe3GeTe2}, where the dense manifold of nearby bands also favors a strong interband response, consistent with the large time-reversal-odd spin accumulation around $+0.4$~eV shown in Fig.~\ref{fig2}(f).

Figure~\ref{fig5} reveals a similar evolution for $\mathbf{m}\parallel+y$. At the Fermi level, Figs.~\ref{fig5}(a) and \ref{fig5}(d) show that the time-reversal-even component contains several Fermi-surface contributions with opposite signs, while the time-reversal-odd response exhibits a pronounced momentum dependence. At the representative hole-side energies, $E-E_F=-0.40$~eV for \ce{Fe3GaTe2} and $-0.33$~eV for \ce{Fe3GeTe2}, the dominant even response is associated mainly with the large $\Gamma$-centered Fermi contours shown in Figs.~\ref{fig5}(b) and \ref{fig5}(e). At the corresponding positive energies of $+0.34$~eV and $+0.41$~eV, the band structures in Figs.~\ref{fig8}(b) and \ref{fig8}(d) show that several Fe-$d$-derived bands approach or cross the selected energies. These bands produce the more complex momentum-resolved distributions displayed in Figs.~\ref{fig5}(c) and \ref{fig5}(f). Notably, rotating the magnetization from $+x$ to $+y$ substantially redistributes the spin accumulation in momentum space, even though the overall band dispersions change only moderately. This contrast highlights the importance of the magnetization-dependent spin character and response matrix elements.

Taken together, Figs.~\ref{fig4}, ~\ref{fig5}, and ~\ref{fig8} show that the pronounced differences between \ce{Fe3GaTe2} and \ce{Fe3GeTe2} originate from their detailed electronic structures rather than from symmetry alone. The time-reversal-even response is governed by the imbalance among Fermi-surface contributions of opposite sign, whereas the time-reversal-odd response is additionally sensitive to interband coupling between nearby electronic states. A related competition among Fermi-surface contributions has been found for monolayer \ce{PtBi2}, where electron- and hole-pocket contributions with opposite signs determine the net Edelstein response~\cite{Pan2026PtBi2}. In the present magnetic systems, this Fermi-surface competition is additionally controlled by the magnetization direction and coexists with the interband time-reversal-odd response. The Ga--Ge substitution therefore modifies both the Fermi-surface geometry and the relative spacing of the Fe-$3d$ bands, giving rise to the distinct magnitude and energy dependence observed in Fig.~~\ref{fig2}.

 \begin{figure*}[htpb]
\centerline{\includegraphics[width=17.2cm]{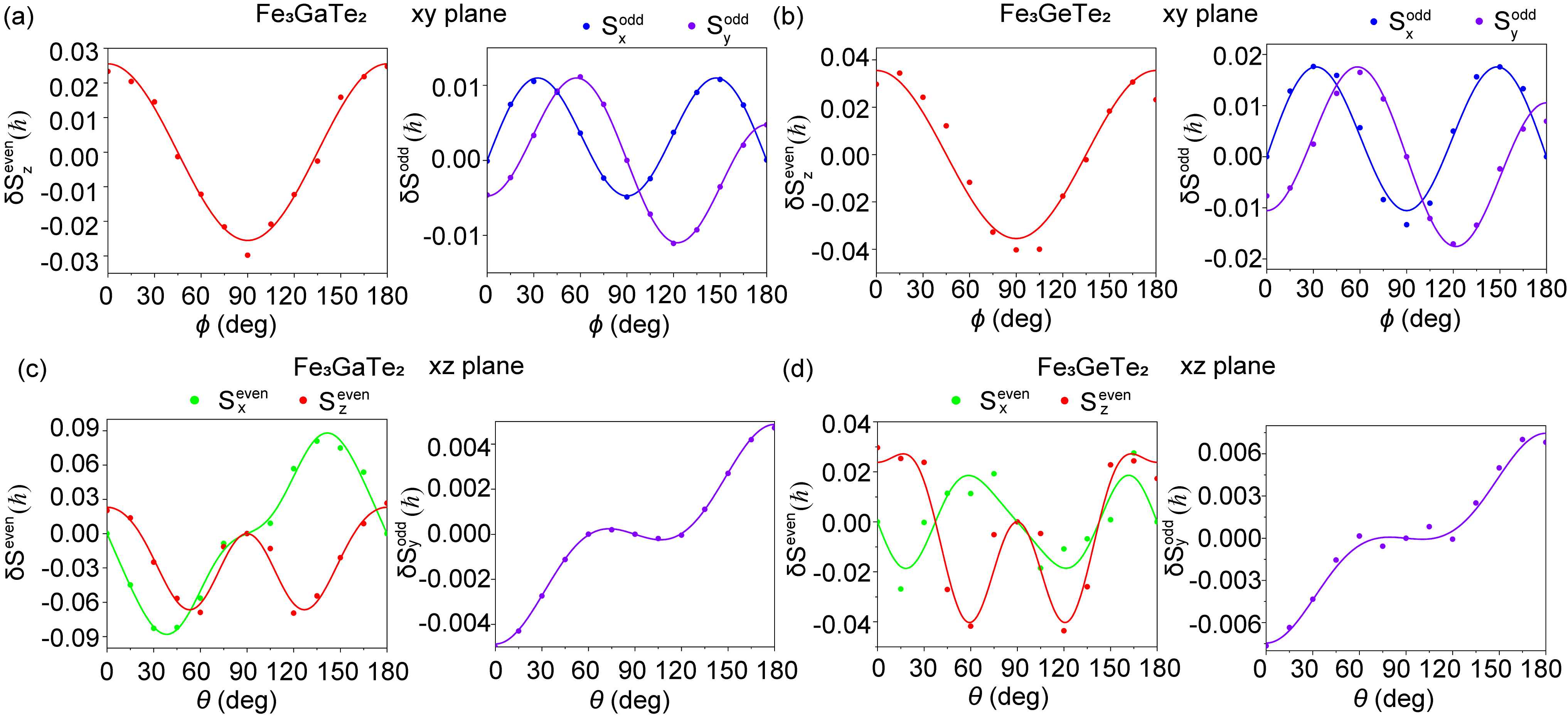}}
\caption{
Angular dependence of the symmetry-allowed current-induced spin accumulation in monolayer \ce{Fe3GaTe2} and \ce{Fe3GeTe2} under an applied electric field of $E_x=1$~V/\AA. (a,b) Results for magnetization rotation in the $xy$ plane for (a) $\mathrm{Fe_3GaTe_2}$ and (b) $\mathrm{Fe_3GeTe_2}$. The time-reversal-even component is fitted by $\delta S_z^{\mathrm{even}}=C\cos(2\phi)$, whereas the time-reversal-odd components are fitted by $\delta S_x^{\mathrm{odd}}=A\sin(3\phi)+B\sin\phi$ and $\delta S_y^{\mathrm{odd}}=-A\cos(3\phi)+B\cos\phi$. (c,d) Corresponding results for magnetization rotation in the $xz$ plane for (c) $\mathrm{Fe_3GaTe_2}$ and (d) $\mathrm{Fe_3GeTe_2}$. The symmetry-allowed responses are fitted by
\(\delta S_x^{\mathrm{even}}=A_2\sin(2\theta)+A_4\sin(4\theta)+A_6\sin(6\theta)\),
\(\delta S_z^{\mathrm{even}}=B_2\cos^2\theta+B_4\cos^4\theta+B_6\cos^6\theta\), and
\(\delta S_y^{\mathrm{odd}}=C_1\cos\theta+C_3\cos^3\theta+C_5\cos^5\theta\).
Here, \(\phi\) and \(\theta\) are measured from the \(+x\) direction for the \(xy\)- and \(xz\)-plane rotations, respectively; \(\phi=90^\circ\) and \(\theta=90^\circ\) correspond to magnetization along \(+y\) and \(+z\), respectively. Symbols denote the calculated values, while solid lines represent the symmetry-constrained fits. The corresponding fitting coefficients are listed in Table~\ref{tab:angular_fit}.
}
\label{fig6}
\end{figure*}

\subsection{Magnetization-angle dependence}

Figure~\ref{fig6} examines the magnetization-direction dependence of the Edelstein response by continuously rotating the magnetization in the $xy$ and $xz$ planes. For rotation in the $xy$ plane, the magnetization is parameterized as $\mathbf{m}=(\cos\phi,\sin\phi,0)$, where $\phi=0^\circ$ and $90^\circ$ correspond to $\mathbf{m}\parallel+x$ and $+y$, respectively. Figures~\ref{fig6}(a) and \ref{fig6}(b) show that, in both \ce{Fe3GaTe2} and \ce{Fe3GeTe2}, only $\delta S_z^{\mathrm{even}}$ is allowed in the even sector, while $\delta S_x^{\mathrm{odd}}$ and $\delta S_y^{\mathrm{odd}}$ remain allowed in the odd sector. The even component follows an approximate $\cos 2\phi$ dependence. The two odd components contain both first- and third-order angular harmonics and are well described by $\delta S_x^{\mathrm{odd}}=A\sin 3\phi+B\sin\phi$ and $\delta S_y^{\mathrm{odd}}=-A\cos 3\phi+B\cos\phi$. Appendix~A provides the detailed symmetry analysis and derivation of these angular dependences. These forms are dictated by the common $D_{3h}$ symmetry and are consistent with the symmetry-adapted expressions previously derived for monolayer 2H-\ce{VTe2} \cite{Pan2026VTe2}.

The parity of these angular dependences directly reflects the underlying magnetic symmetry. The transformation $\phi\rightarrow180^\circ-\phi$ is equivalent to the $M_xT$ operation, under which $\delta S_z^{\mathrm{even}}$ and $\delta S_x^{\mathrm{odd}}$ are symmetric about $\phi=90^\circ$, whereas $\delta S_y^{\mathrm{odd}}$ is antisymmetric. Although these parity relations are identical in the two compounds, their response amplitudes differ quantitatively. In \ce{Fe3GaTe2}, $\delta S_z^{\mathrm{even}}$ varies from approximately $+0.025\,\hbar$ at $\phi=0^\circ$ and $180^\circ$ to $-0.025\,\hbar$ near $\phi=90^\circ$, whereas the corresponding extrema in \ce{Fe3GeTe2} are approximately $+0.035\,\hbar$ and $-0.035\,\hbar$. The odd responses are also stronger in \ce{Fe3GeTe2}: $\delta S_x^{\mathrm{odd}}$ spans approximately $-0.011$ to $+0.018\,\hbar$ and $\delta S_y^{\mathrm{odd}}$ spans approximately $-0.018$ to $+0.018\,\hbar$, compared with respective ranges of about $-0.006$ to $+0.011\,\hbar$ and $-0.012$ to $+0.011\,\hbar$ in \ce{Fe3GaTe2}. Thus, while the common $D_{3h}$ symmetry determines the allowed components and their parity relations, \ce{Fe3GeTe2} exhibits systematically larger $xy$-plane Edelstein responses, reflecting the material dependence of their underlying electronic structures.

Figures~\ref{fig6}(c) and \ref{fig6}(d) reveal a richer angular dependence for magnetization rotation in the $xz$ plane. In this geometry, the magnetization is parameterized as $\mathbf{m}=(\cos\theta,0,\sin\theta)$, with $\theta$ measured from the $+x$ direction. In this geometry, $\delta S_x^{\mathrm{even}}$, $\delta S_z^{\mathrm{even}}$, and $\delta S_y^{\mathrm{odd}}$ are symmetry allowed. At the lowest order, these components follow $\sin 2\theta$, $\cos^2\theta$, and $\cos\theta$, respectively, consistent with the lowest-order $D_{3h}$ symmetry expansion of current-induced spin polarization \cite{Li2020PNAS}. The first-principles results, however, reveal appreciable higher-order angular contributions, particularly in the even sector. They are well described by $\delta S_x^{\mathrm{even}}=A_2\sin 2\theta+A_4\sin 4\theta+A_6\sin 6\theta$, $\delta S_z^{\mathrm{even}}=B_2\cos^2\theta+B_4\cos^4\theta+B_6\cos^6\theta$, and $\delta S_y^{\mathrm{odd}}=C_1\cos\theta+C_3\cos^3\theta+C_5\cos^5\theta$. The symmetry derivation of these expressions is given in Appendix~\ref{app:angular}, while the corresponding fitting coefficients are listed in Table~II. Quantitatively, the two compounds exhibit distinct component-dependent response amplitudes.  \ce{Fe3GaTe2} exhibits substantially larger time-reversal-even responses in the $xz$-rotation geometry. The maximum magnitude of $\delta S_x^{\mathrm{even}}$ reaches approximately $0.083\hbar$ in \ce{Fe3GaTe2}, more than three times the value of about $0.028\hbar$ in \ce{Fe3GeTe2}. Likewise, the maximum magnitude of $\delta S_z^{\mathrm{even}}$ is approximately $0.069\hbar$ in \ce{Fe3GaTe2}, compared with $0.040\hbar$ in \ce{Fe3GeTe2}. By contrast, the time-reversal-odd response is stronger in \ce{Fe3GeTe2}, where $\delta S_y^{\mathrm{odd}}$ reaches approximately $7.5\times10^{-3},\hbar$, compared with $4.9\times10^{-3},\hbar$ in \ce{Fe3GaTe2}. Despite its smaller overall $\delta S_x^{\mathrm{even}}$ amplitude, \ce{Fe3GeTe2} exhibits a more complex angular profile with additional extrema and sign changes, indicating a larger relative contribution from higher-order angular harmonics.

The $xz$ rotation also exhibits characteristic parity constraints. Under $\theta\rightarrow180^\circ-\theta$, $\delta S_x^{\mathrm{even}}$ and $\delta S_y^{\mathrm{odd}}$ change sign, whereas $\delta S_z^{\mathrm{even}}$ remains unchanged. All three components vanish at $\theta=90^\circ$, corresponding to $\mathbf{m}\parallel z$, where the threefold rotational symmetry is restored. Although these symmetry constraints are common to both materials, the detailed angular profiles differ substantially, reflecting their different electronic structures and higher-order response coefficients. The comparison therefore demonstrates that crystal symmetry determines the allowed components and their parity relations, whereas the electronic structure controls their magnitudes and higher-harmonic content. Notably, substantial higher-order angular harmonics have also been identified in first-principles calculations of the intrinsic SOT in monolayer \ce{Fe3GeTe2}, where they were shown to play an essential role in deterministic field-free magnetization switching~\cite{Xue2023FGT}. The present results show that such nontrivial angular structure already emerges at the level of the underlying nonequilibrium spin accumulation, providing a microscopic connection between the angular dependence of the Edelstein response and that of the current-induced torque.

\begin{table}[t]
\caption{
Fitting coefficients for the angular-dependent current-induced spin
accumulation of monolayer $\mathrm{Fe_3GaTe_2}$ and
$\mathrm{Fe_3GeTe_2}$ shown in Fig. 6.
The coefficients correspond to the symmetry-constrained fitting
functions given in the figure caption. All coefficients are in units
of $\hbar$.
}
\label{tab:angular_fit}
\centering
\begin{tabular}{c c c}
\hline\hline
Coefficient
& $\mathrm{Fe_3GaTe_2}$
& $\mathrm{Fe_3GeTe_2}$ \\
\hline

\multicolumn{3}{c}{$xy$-plane magnetization rotation} \\
\hline
$C$   & $0.02506$ & $0.03548$ \\
$A$   & $0.008875$ & $0.01518$ \\
$B$   & $0.004077$ & $0.004646$ \\

\hline
\multicolumn{3}{c}{$xz$-plane magnetization rotation} \\
\hline
$A_2$ & $-7.597\times10^{-2}$ & $ 4.676\times10^{-3}$ \\
$A_4$ & $-1.808\times10^{-2}$ & $-1.666\times10^{-2}$ \\
$A_6$ & $ 7.802\times10^{-3}$ & $-5.796\times10^{-3}$ \\

$B_2$ & $-4.072\times10^{-1}$ & $-3.410\times10^{-1}$ \\
$B_4$ & $ 7.239\times10^{-1}$ & $ 8.394\times10^{-1}$ \\
$B_6$ & $-2.936\times10^{-1}$ & $-4.745\times10^{-1}$ \\

$C_1$ & $ 1.279\times10^{-3}$ & $ 6.078\times10^{-4}$ \\
$C_3$ & $-5.130\times10^{-3}$ & $-6.055\times10^{-3}$ \\
$C_5$ & $-1.020\times10^{-3}$ & $-2.010\times10^{-3}$ \\
\hline\hline
\end{tabular}
\end{table}

\begin{figure*}[htpb]
\includegraphics[width=17.2cm]{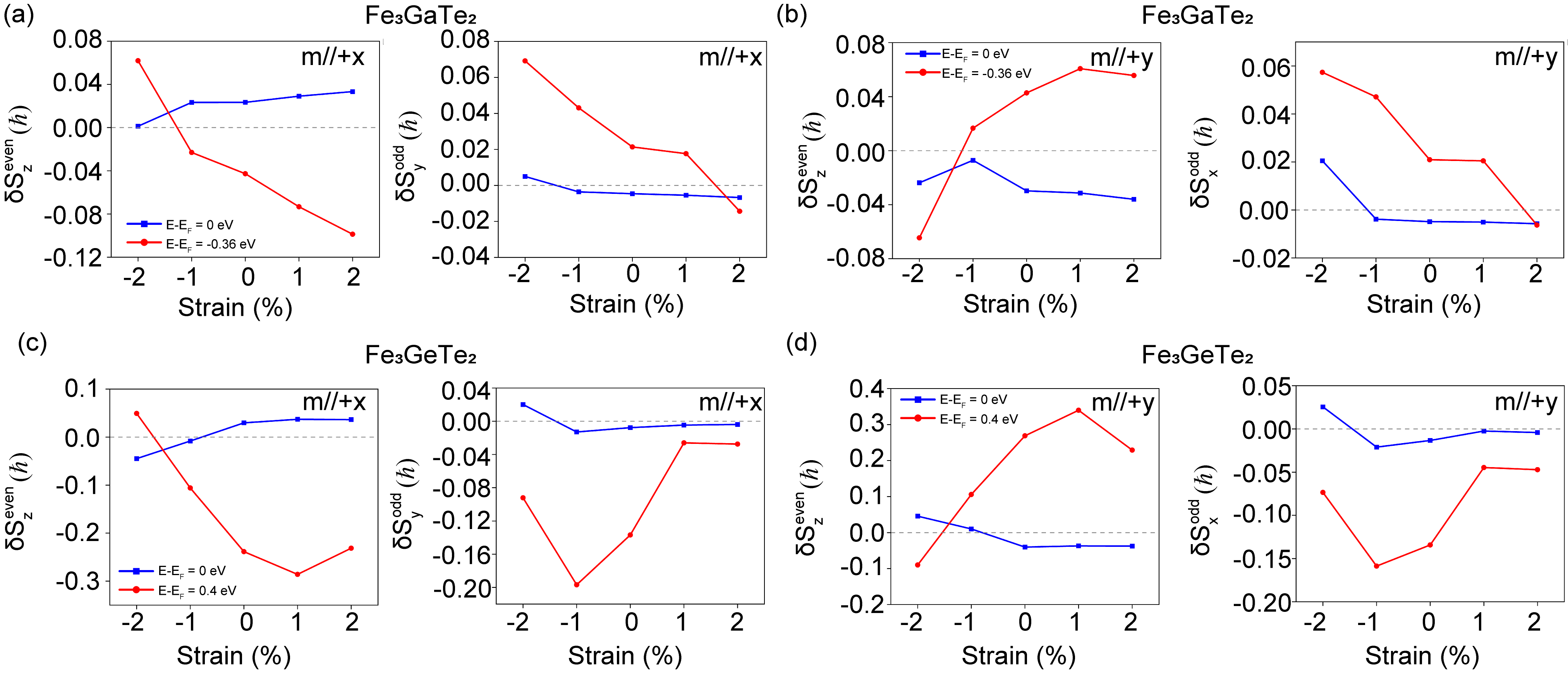}
\caption{ Strain dependence of the current-induced spin accumulations at selected characteristic energies in monolayer \ce{Fe3GaTe2} and \ce{Fe3GeTe2}, under an applied electric field of $E_x=1$~V/\AA. The data points are extracted from the energy-dependent Edelstein response shown in Fig.~\ref{fig10}. 
Panels (a) and (b) show the results for \ce{Fe3GaTe2} with magnetization along the $+x$ and $+y$ directions, respectively, while panels (c) and (d) show the corresponding results for \ce{Fe3GeTe2}.
For $\mathbf{m}\parallel +x$, the time-reversal-even $\delta S_{z}^{\mathrm{even}}$ and time-reversal-odd $\delta S_{y}^{\mathrm{odd}}$ components are displayed, whereas for $\mathbf{m}\parallel +y$, $\delta S_{z}^{\mathrm{even}}$ and $\delta S_{x}^{\mathrm{odd}}$ are shown.
The blue curves correspond to the intrinsic Fermi level ($E-E_{\mathrm F}=0$), while the red curves represent selected energies away from the Fermi level: $E-E_{\mathrm F}=-0.36$~eV for \ce{Fe3GaTe2} and $+0.40$~eV for \ce{Fe3GeTe2}.
The biaxial strain ranges from $-2\%$ to $+2\%$, with negative and positive values denoting compressive and tensile strain, respectively.}
\label{fig7}
\end{figure*}

\subsection{Strain tunability of the Edelstein response}

We next examine the strain dependence of the Edelstein response. Since monolayer materials are commonly synthesized or transferred onto substrates in experiments, lattice mismatch between the monolayer and substrate can introduce residual biaxial strain. Such strain may modify the electronic states near the Fermi level and therefore affect the current-induced spin accumulation. Figure~\ref{fig7} summarizes the symmetry-allowed spin accumulations under different biaxial strains at the intrinsic Fermi level and at representative energies where pronounced responses are observed in the unstrained systems. The corresponding energy-dependent Edelstein responses for strains ranging from $-2\%$ to $+2\%$ are presented in Fig.~\ref{fig10} of the Appendix. Similar strain-dependent variations of the Edelstein response have been reported in monolayer \ce{PtBi2}, where the changes were attributed to strain-induced modifications of the Fermi-surface electronic states~\cite{Pan2026PtBi2}.

For \ce{Fe3GaTe2}, Figs.~\ref{fig7}(a) and \ref{fig7}(b) show that the response at the intrinsic Fermi level varies relatively weakly with strain. For $\mathbf{m}\parallel+x$, $\delta S_z^{\mathrm{even}}$ increases gradually under tensile strain, whereas $\delta S_y^{\mathrm{odd}}$ remains close to zero. A much stronger variation appears at $E-E_F=-0.36$~eV. As shown in Fig.~\ref{fig7}(a), $\delta S_z^{\mathrm{even}}$ changes from positive under $-2\%$ compressive strain to increasingly negative values under tensile strain, while $\delta S_y^{\mathrm{odd}}$ decreases from approximately $0.07\,\hbar$ to a small negative value. For $\mathbf{m}\parallel+y$, Fig.~\ref{fig7}(b) shows that $\delta S_z^{\mathrm{even}}$ at the same energy changes from negative at $-2\%$ strain to positive above approximately $-1\%$. Meanwhile, $\delta S_x^{\mathrm{odd}}$ decreases continuously and eventually changes sign near $+2\%$. Figures~\ref{fig7}(c) and \ref{fig7}(d) reveal an even stronger strain dependence in \ce{Fe3GeTe2}, particularly at $E-E_F=+0.40$~eV. For $\mathbf{m}\parallel+x$, Fig.~\ref{fig7}(c) shows that $\delta S_z^{\mathrm{even}}$ evolves from a small positive value at $-2\%$ strain to a large negative response approaching $-0.3\,\hbar$ near $+1\%$. The corresponding $\delta S_y^{\mathrm{odd}}$ varies nonmonotonically and reaches its largest magnitude near $-1\%$ strain. For $\mathbf{m}\parallel+y$, Fig.~\ref{fig7}(d) shows an even more pronounced sign reversal of the even component, which increases from approximately $-0.1\,\hbar$ at $-2\%$ strain to above $0.3\,\hbar$ at $+1\%$. The odd component remains negative but varies substantially in magnitude. The full spectra in Fig.~\ref{fig10} further demonstrate that strain shifts and reshapes the characteristic peaks and sign-changing features rather than simply rescaling the response. This strong tunability is consistent with the broader sensitivity of current-induced responses in \ce{Fe3GeTe2}-based systems to electronic-structure engineering. Carrier filling can substantially modify the balance among current-induced torque contributions in bulk \ce{Fe3GeTe2} \cite{Saunderson2022FGT}, while ferroelectric proximity can reconstruct the electronic structure and strongly modulate the torkance in \ce{Fe3GeTe2}/\ce{In2Se3} \cite{Pan2026FGTIn2Se3}. In contrast to these approaches, biaxial strain provides a direct single-layer route for tuning the Edelstein spin accumulation without introducing an additional interface.


\section{CONCLUSION} 

In summary, we have systematically investigated the Edelstein response of monolayer \ce{Fe3GaTe2} and \ce{Fe3GeTe2} using first-principles Wannier-based linear-response calculations. Although the two compounds share the same noncentrosymmetric $D_{3h}$ crystal symmetry and consequently obey the same symmetry selection rules, they exhibit markedly different current-induced spin accumulations. For an electric field applied along $+x$, $\delta S_z^{\mathrm{even}}$ and $\delta S_y^{\mathrm{odd}}$ are allowed for $\mathbf{m}\parallel+x$, whereas $\delta S_z^{\mathrm{even}}$ and $\delta S_x^{\mathrm{odd}}$ survive for $\mathbf{m}\parallel+y$. Their magnitudes, signs, and characteristic energy scales, however, depend strongly on the underlying electronic structure. Pronounced responses in \ce{Fe3GaTe2} occur mainly on the hole side around $-0.4$~eV, whereas the largest responses in \ce{Fe3GeTe2} appear predominantly on the electron side around $+0.4$~eV.

At the intrinsic Fermi level, the atom-resolved decomposition reveals substantial cancellation among different atomic contributions and shows that replacing Ga with Ge reverses the even contribution of the inequivalent Fe3 site and strengthens its negative odd contribution. The momentum-resolved calculations at selected energies further demonstrate that the even response reflects an incomplete cancellation among opposite-sign intraband Fermi-surface contributions, whereas the odd response is enhanced by interband coherence between nearby Fe-$3d$ states. Their contrasting broadening dependences support this distinction: the even response displays the characteristic intraband enhancement with decreasing broadening, while the odd response remains much less sensitive to broadening, consistent with its interband origin.

The magnetization-direction dependence provides an additional manifestation of the interplay between symmetry and electronic structure. For rotation in the $xy$ plane, the calculated responses follow the symmetry-constrained angular harmonics imposed by the $D_{3h}$ crystal symmetry. Rotation in the $xz$ plane reveals substantial higher-order contributions beyond the lowest-order $\sin2\theta$, $\cos^2\theta$, and $\cos\theta$ forms. In this geometry, \ce{Fe3GaTe2} exhibits substantially larger time-reversal-even responses, whereas \ce{Fe3GeTe2} shows a stronger time-reversal-odd response and a comparatively more pronounced higher-harmonic structure in $\delta S_x^{\mathrm{even}}$. All symmetry-allowed components vanish for $\mathbf{m}\parallel z$ when the threefold rotational symmetry is restored.

Finally, biaxial strain strongly modifies the Edelstein response away from the intrinsic Fermi level, producing pronounced nonmonotonic variations and, in several cases, sign reversals. These results establish crystal symmetry, electronic structure, magnetization orientation, and lattice deformation as complementary control parameters for the Edelstein effect in the \ce{Fe3$X$Te2} ($X=\mathrm{Ga},\mathrm{Ge}$) family and provide a microscopic basis for tailoring the spin-accumulation channel underlying current-induced torques in two-dimensional ferromagnets.

\section*{ACKNOWLEDGMENTS}

This work was financially supported by  the  China Postdoctoral Science Foundation under Grant Number GZC20261825.


\appendix

\section{Symmetry-constrained angular dependence of the Edelstein effect}
\label{app:angular}

\begin{figure}[!htpb]
\includegraphics[width=8cm]{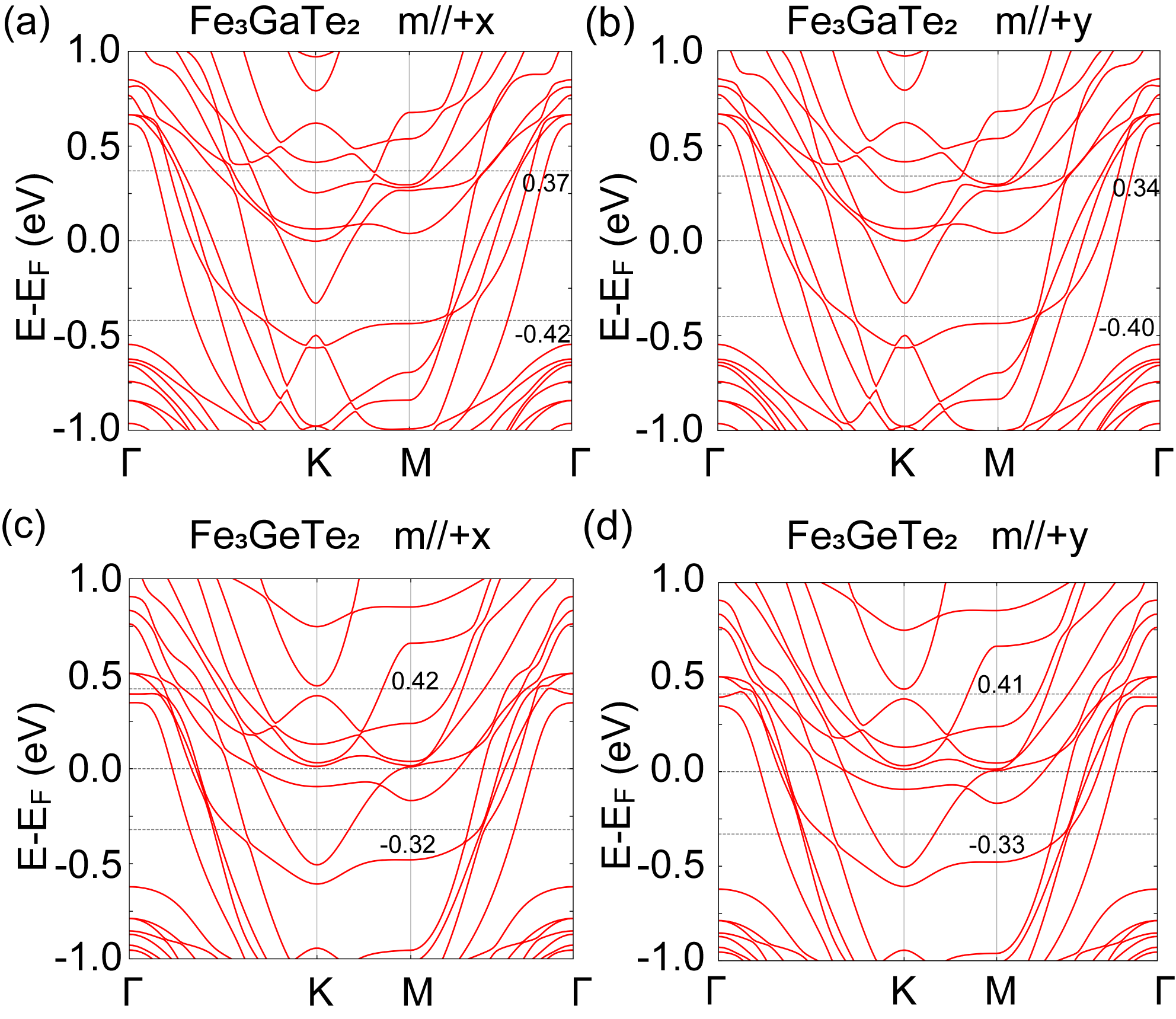}
\caption{Calculated band structures of monolayer \ce{Fe3GaTe2} and \ce{Fe3GeTe2} for different in-plane magnetization directions. Panels (a) and (b) show the band structures of \ce{Fe3GaTe2} for magnetization along the $+x$ and $+y$ directions, respectively, whereas panels (c) and (d) show the corresponding results for \ce{Fe3GeTe2}. The labeled energies indicate the representative energy positions selected from the extrema of the energy-dependent current-induced spin accumulations in Fig.~\ref{fig2} and used for the momentum-resolved analysis in Figs.~\ref{fig4} and ~\ref{fig5}.}
\label{fig8}
\end{figure}

To understand the magnetization-direction dependence of the current-induced spin accumulation in monolayer \ce{Fe3GaTe2} and \ce{Fe3GeTe2}, we employ a symmetry-based expansion of the Edelstein response. Both monolayers belong to the $D_{3h}$ point group, whose relevant generators may be chosen as $C_{3z}$, $M_x$, and $M_z$. The current-induced spin accumulation is written as
\begin{equation}
    \delta S_i
    =
    \sum_{j=x,y}
    \chi_{ij}(\mathbf{m}) E_j ,
\label{eq:app_response}
\end{equation}
where $\mathbf{m}=\mathbf{M}/|\mathbf{M}|$ denotes the unit vector along the
magnetization direction. Throughout the present work, the electric field is
applied along the $+x$ direction, $\mathbf{E}=E_x\hat{\mathbf{x}}$.

For a magnetic system, the Edelstein susceptibility can be decomposed into
time-reversal-even and time-reversal-odd parts. These satisfy
$\chi_{ij}^{\mathrm{even}}(-\mathbf{m})
=\chi_{ij}^{\mathrm{even}}(\mathbf{m})$ and
$\chi_{ij}^{\mathrm{odd}}(-\mathbf{m})
=-\chi_{ij}^{\mathrm{odd}}(\mathbf{m})$. Accordingly, the even response
contains only even powers of the magnetization direction, whereas the odd
response contains only odd powers. Following the symmetry-expansion approach
used for Edelstein responses~\cite{Pan2026VTe2,Li2020PNAS}, we write
\begin{equation}
\begin{aligned}
\chi_{ij}^{\mathrm{even}}
=&\,
\chi_{ij}^{(0)}
+
\sum_{ab}
\chi_{ij;ab}^{(2)}m_am_b
+
\sum_{abcd}
\chi_{ij;abcd}^{(4)}m_am_bm_cm_d
\\
&+
\sum_{abcdef}
\chi_{ij;abcdef}^{(6)}
m_am_bm_cm_dm_em_f
+
\mathcal{O}(m^8),
\\[2pt]
\chi_{ij}^{\mathrm{odd}}
=&\,
\sum_a
\chi_{ij;a}^{(1)}m_a
+
\sum_{abc}
\chi_{ij;abc}^{(3)}m_am_bm_c
\\
&+
\sum_{abcde}
\chi_{ij;abcde}^{(5)}
m_am_bm_cm_dm_e
+
\mathcal{O}(m^7).
\end{aligned}
\label{eq:general_expansion}
\end{equation}
Here $a,b,\ldots=x,y,z$. Thus, when terms up to sixth order in the
magnetization direction are retained, the even response contains terms through
$m^6$, whereas the odd response contains terms through $m^5$.

Under the mirror operation $M_x$, the relevant quantities transform as
$(E_x,E_y)\rightarrow(-E_x,E_y)$,
$(m_x,m_y,m_z)\rightarrow(m_x,-m_y,-m_z)$, and
$(S_x,S_y,S_z)\rightarrow(S_x,-S_y,-S_z)$. Under $M_z$, they transform as
$(E_x,E_y)\rightarrow(E_x,E_y)$,
$(m_x,m_y,m_z)\rightarrow(-m_x,-m_y,m_z)$, and
$(S_x,S_y,S_z)\rightarrow(-S_x,-S_y,S_z)$. Here $\mathbf{E}$ is a polar
vector, whereas both $\mathbf{m}$ and $\mathbf{S}$ are axial vectors. Under
$C_{3z}$, the in-plane components of $\mathbf{E}$, $\mathbf{m}$, and
$\mathbf{S}$ transform simultaneously according to the corresponding
two-dimensional rotation matrix.

At the lowest nonvanishing order, imposing the constraints of $C_{3z}$,
$M_x$, and $M_z$ gives
\begin{equation}
\begin{aligned}
\chi^{\mathrm{even},(2)}
&=
\begin{pmatrix}
c_1m_xm_z        & -c_1m_ym_z \\
-c_1m_ym_z       & -c_1m_xm_z \\
c_2(m_x^2-m_y^2) & -2c_2m_xm_y
\end{pmatrix},
\\[4pt]
\chi^{\mathrm{odd},(1)}
&=
\begin{pmatrix}
c_3m_y & c_3m_x \\
c_3m_x & -c_3m_y \\
0      & 0
\end{pmatrix}.
\end{aligned}
\label{eq:lowest_tensors}
\end{equation}
These expressions reproduce the lowest-order current-induced spin
polarizations previously derived for magnetic monolayers with $D_{3h}$
symmetry~\cite{Li2020PNAS,Pan2026VTe2}. They provide a common starting point
for analyzing magnetization rotation within both the $xy$ and $xz$ planes.

\subsection{Magnetization rotation in the $xy$ plane}
\label{app:xy_rotation}

We first consider an in-plane rotation of the magnetization,
$\mathbf{m}(\phi)=(\cos\phi,\sin\phi,0)$, where $\phi$ is measured from the
$+x$ direction. Thus, $\phi=0^\circ$ and $90^\circ$ correspond to
magnetization along $+x$ and $+y$, respectively. For an electric field
applied along $+x$, substitution into Eq.~(\ref{eq:lowest_tensors}) gives the
lowest-order responses
\begin{equation}
\begin{aligned}
\frac{\delta S_z^{\mathrm{even}}}{E_x}
&=
c_2\cos 2\phi,
\\
\frac{\delta S_x^{\mathrm{odd}}}{E_x}
&=
c_3\sin\phi,
\\
\frac{\delta S_y^{\mathrm{odd}}}{E_x}
&=
c_3\cos\phi .
\end{aligned}
\label{eq:xy_lowest}
\end{equation}
Thus, the time-reversal-even response starts from the second angular harmonic,
whereas the time-reversal-odd response contains the fundamental harmonic.

Higher-order odd terms allowed by the same $D_{3h}$ symmetry generate an
additional third angular harmonic. Retaining the leading symmetry-allowed
higher-order contribution gives
\begin{equation}
\begin{aligned}
\frac{\delta S_z^{\mathrm{even}}}{E_x}
&=
C\cos 2\phi,
\\
\frac{\delta S_x^{\mathrm{odd}}}{E_x}
&=
A\sin 3\phi+B\sin\phi,
\\
\frac{\delta S_y^{\mathrm{odd}}}{E_x}
&=
-A\cos 3\phi+B\cos\phi .
\end{aligned}
\label{eq:xy_fit}
\end{equation}
These symmetry-adapted angular forms are identical to those previously
obtained for monolayer $2H$-\ce{VTe2} with $D_{3h}$
symmetry~\cite{Pan2026VTe2}, and are used to describe the calculated
$xy$-plane angular dependences in Fig.~\ref{fig6}.

The parity about $\phi=90^\circ$ follows directly from symmetry. The
transformation $\phi\rightarrow\pi-\phi$, corresponding to
$(m_x,m_y)\rightarrow(-m_x,m_y)$, is related to the $M_xT$ operation for the
in-plane magnetic configuration. Consequently,
$\delta S_z^{\mathrm{even}}(\pi-\phi)=
\delta S_z^{\mathrm{even}}(\phi)$ and
$\delta S_x^{\mathrm{odd}}(\pi-\phi)=
\delta S_x^{\mathrm{odd}}(\phi)$, whereas
$\delta S_y^{\mathrm{odd}}(\pi-\phi)=
-\delta S_y^{\mathrm{odd}}(\phi)$. These relations account for the symmetric
behavior of $\delta S_z^{\mathrm{even}}$ and $\delta S_x^{\mathrm{odd}}$, and
the antisymmetric behavior of $\delta S_y^{\mathrm{odd}}$ about
$\phi=90^\circ$ observed in Fig.~~\ref{fig6}.

\subsection{Magnetization rotation in the $xz$ plane}
\label{app:xz_rotation}

We next consider rotation of the magnetization in the $xz$ plane. The angle
$\theta$ is defined from the $+x$ direction through
$\mathbf{m}(\theta)=(\cos\theta,0,\sin\theta)$, such that
$\theta=0^\circ$, $90^\circ$, and $180^\circ$ correspond to magnetization
along $+x$, $+z$, and $-x$, respectively.

Substitution into Eq.~(\ref{eq:lowest_tensors}) gives
\begin{equation}
\begin{aligned}
\chi_{xz}^{\mathrm{even},(2)}(\theta)
&=
\begin{pmatrix}
c_1\cos\theta\sin\theta & 0 \\
0 & -c_1\cos\theta\sin\theta \\
c_2\cos^2\theta & 0
\end{pmatrix},
\\[6pt]
\chi_{xz}^{\mathrm{odd},(1)}(\theta)
&=
\begin{pmatrix}
0 & c_3\cos\theta \\
c_3\cos\theta & 0 \\
0 & 0
\end{pmatrix}.
\end{aligned}
\label{eq:xz_lowest_tensors}
\end{equation}
For $\mathbf{E}=E_x\hat{\mathbf{x}}$, the corresponding lowest-order spin
accumulations are
\begin{equation}
\begin{aligned}
\delta S_x^{\mathrm{even}}
&=
\frac{c_1}{2}E_x\sin2\theta,
\\
\delta S_z^{\mathrm{even}}
&=
c_2E_x\cos^2\theta,
\\
\delta S_y^{\mathrm{odd}}
&=
c_3E_x\cos\theta,
\end{aligned}
\label{eq:xz_lowest_spin}
\end{equation}
while $\delta S_y^{\mathrm{even}}=\delta S_x^{\mathrm{odd}}
=\delta S_z^{\mathrm{odd}}=0$. Thus, canting the magnetization out of the
plane activates an additional longitudinal time-reversal-even component
$\delta S_x^{\mathrm{even}}$, which vanishes for both purely in-plane and
purely out-of-plane magnetization.

To construct the higher-order expansion, we first identify the parity
constraints on the symmetry-allowed tensor elements. For $m_y=0$, $M_z$ maps
$(m_x,m_z)\rightarrow(-m_x,m_z)$, whereas $M_x$ maps
$(m_x,m_z)\rightarrow(m_x,-m_z)$. Since $E_x$ is even under $M_z$ and odd
under $M_x$, the allowed tensor components satisfy
\begin{equation}
\begin{aligned}
\chi_{xx}(-m_x,m_z) &= -\chi_{xx}(m_x,m_z),\\
\chi_{xx}(m_x,-m_z) &= -\chi_{xx}(m_x,m_z),\\
\chi_{zx}(-m_x,m_z) &= +\chi_{zx}(m_x,m_z),\\
\chi_{zx}(m_x,-m_z) &= +\chi_{zx}(m_x,m_z),\\
\chi_{yx}(-m_x,m_z) &= -\chi_{yx}(m_x,m_z),\\
\chi_{yx}(m_x,-m_z) &= +\chi_{yx}(m_x,m_z).
\end{aligned}
\label{eq:parity_xz}
\end{equation}
Combining these spatial constraints with the even/odd behavior under
$\mathbf{m}\rightarrow-\mathbf{m}$ shows that, for an electric field along
$x$, $\chi_{xx}$ and $\chi_{zx}$ belong exclusively to the
time-reversal-even sector, whereas $\chi_{yx}$ belongs exclusively to the
time-reversal-odd sector. Hence
$\chi_{yx}^{\mathrm{even}}=\chi_{xx}^{\mathrm{odd}}
=\chi_{zx}^{\mathrm{odd}}=0$.

The allowed monomials can then be constructed order by order. For
$\chi_{xx}^{\mathrm{even}}$, both the powers of $m_x$ and $m_z$ must be odd:
$m_xm_z$ at second order; $m_x^3m_z$ and $m_xm_z^3$ at fourth order; and
$m_x^5m_z$, $m_x^3m_z^3$, and $m_xm_z^5$ at sixth order. For
$\chi_{zx}^{\mathrm{even}}$, both powers must be even and the response must
vanish for $m_x=0$, giving $m_x^2$ at second order;
$m_x^4$ and $m_x^2m_z^2$ at fourth order; and
$m_x^6$, $m_x^4m_z^2$, and $m_x^2m_z^4$ at sixth order. Finally, for
$\chi_{yx}^{\mathrm{odd}}$, the power of $m_x$ must be odd and that of $m_z$
must be even, giving $m_x$ at first order;
$m_x^3$ and $m_xm_z^2$ at third order; and
$m_x^5$, $m_x^3m_z^2$, and $m_xm_z^4$ at fifth order.

\begin{figure}[!htpb]
\includegraphics[width=8.5cm]{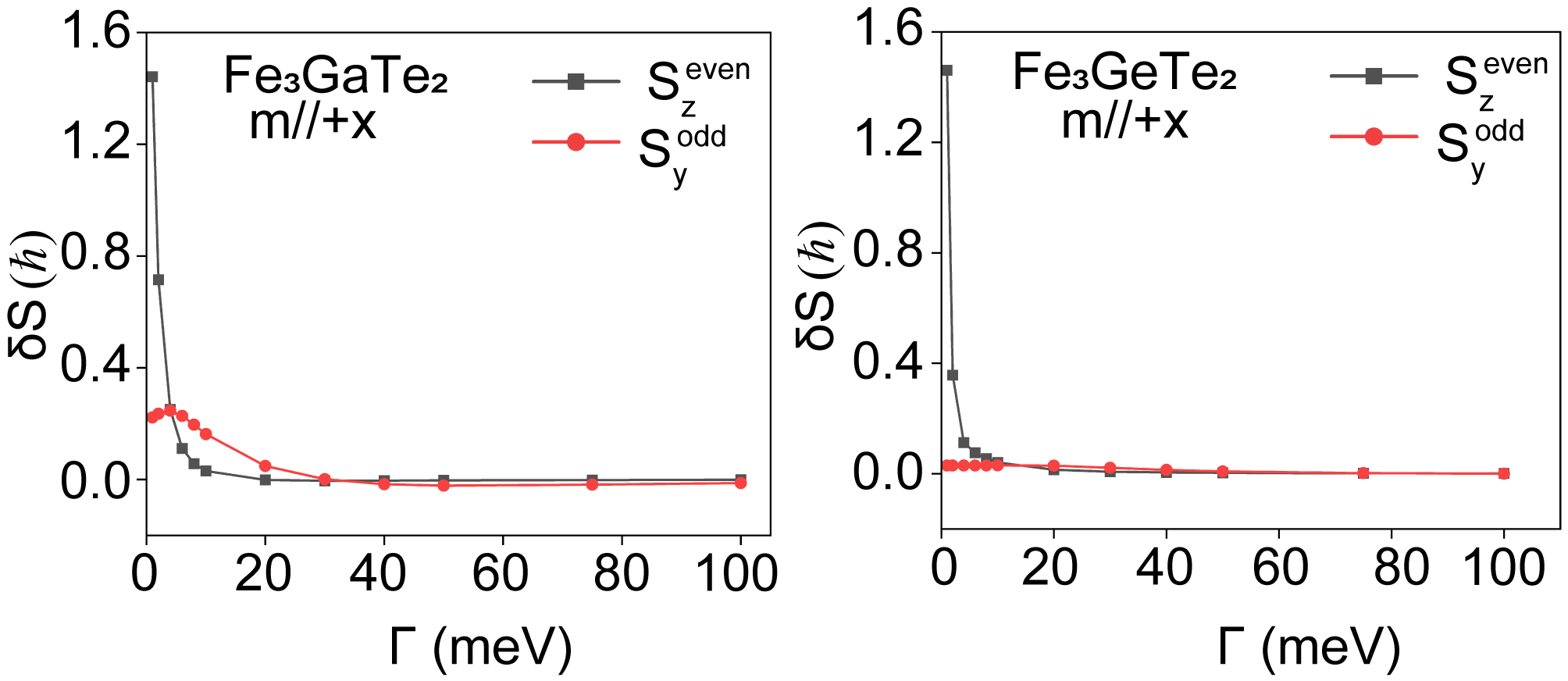}
\caption{Broadening dependence of the current-induced spin accumulation in monolayer (a) \ce{Fe3GaTe2} and (b) \ce{Fe3GeTe2} for $\mathbf{m}\parallel+x$ under a reference electric field of $1$~V/\AA\ applied along $+x$. The energy is fixed at $E-E_F=-0.40$~eV. The time-reversal-even $\delta S_z^{\mathrm{even}}$ and time-reversal-odd $\delta S_y^{\mathrm{odd}}$ components are shown as functions of the phenomenological broadening $\Gamma$.}
\label{fig9}
\end{figure}

\begin{figure*}[!htpb]
\includegraphics[width=17.2cm]{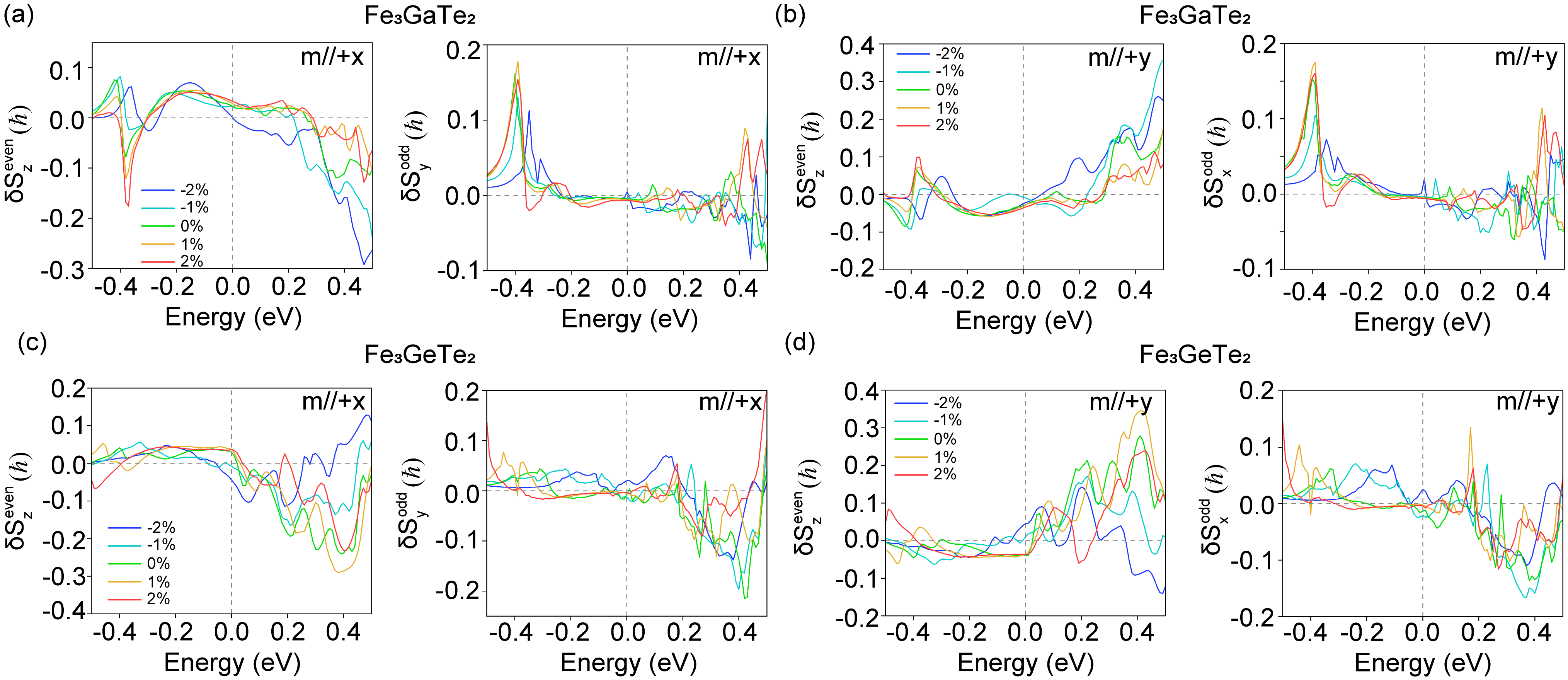}
\caption{ Strain-dependent current-induced spin accumulations as functions of Fermi energy in monolayer \ce{Fe3GaTe2} and \ce{Fe3GeTe2} under a reference electric field of 1~V/\AA\ applied along the $+x$ direction. Panels (a) and (b) show the results for \ce{Fe3GaTe2} with magnetization along the $+x$ and $+y$ directions, respectively, whereas panels (c) and (d) show the corresponding results for \ce{Fe3GeTe2}. For $\mathbf{m}\parallel +x$, the time-reversal-even $\delta S_{z}^{\mathrm{even}}$ and time-reversal-odd $\delta S_{y}^{\mathrm{odd}}$ components are shown, while for $\mathbf{m}\parallel +y$, the corresponding $\delta S_{z}^{\mathrm{even}}$ and $\delta S_{x}^{\mathrm{odd}}$ components are displayed. The biaxial strain is varied from $-2\%$ to $+2\%$, with negative and positive values denoting compressive and tensile strain, respectively.}
\label{fig10}
\end{figure*}

Since $\mathbf{m}$ is a unit vector, $m_x^2+m_z^2=1$, these monomials are
not all linearly independent along the one-dimensional $xz$ rotation path.
A minimal nonredundant polynomial basis can therefore be written as
\begin{equation}
\begin{aligned}
\chi_{xx}^{\mathrm{even}}
&=
m_xm_z
\left(
\alpha_2+\alpha_4m_z^2+\alpha_6m_z^4
\right),
\\
\chi_{zx}^{\mathrm{even}}
&=
m_x^2
\left(
\beta_2+\beta_4m_z^2+\beta_6m_z^4
\right),
\\
\chi_{yx}^{\mathrm{odd}}
&=
m_x
\left(
\gamma_1+\gamma_3m_z^2+\gamma_5m_z^4
\right).
\end{aligned}
\label{eq:xz_poly}
\end{equation}
Using $m_x=\cos\theta$ and $m_z=\sin\theta$, these expressions become
\begin{equation}
\begin{aligned}
\frac{\delta S_x^{\mathrm{even}}}{E_x}
&=
\cos\theta\sin\theta
\left(
\alpha_2+\alpha_4\sin^2\theta+\alpha_6\sin^4\theta
\right),
\\
\frac{\delta S_z^{\mathrm{even}}}{E_x}
&=
\cos^2\theta
\left(
\beta_2+\beta_4\sin^2\theta+\beta_6\sin^4\theta
\right),
\\
\frac{\delta S_y^{\mathrm{odd}}}{E_x}
&=
\cos\theta
\left(
\gamma_1+\gamma_3\sin^2\theta+\gamma_5\sin^4\theta
\right).
\end{aligned}
\label{eq:xz_angular_poly}
\end{equation}
These forms automatically satisfy
$\delta S_x^{\mathrm{even}}(\pi-\theta)=
-\delta S_x^{\mathrm{even}}(\theta)$,
$\delta S_z^{\mathrm{even}}(\pi-\theta)=
+\delta S_z^{\mathrm{even}}(\theta)$, and
$\delta S_y^{\mathrm{odd}}(\pi-\theta)=
-\delta S_y^{\mathrm{odd}}(\theta)$. At $\theta=\pi/2$, corresponding to
$\mathbf{m}\parallel z$, all three components vanish as required by the
restored threefold rotational symmetry. In addition,
$\delta S_x^{\mathrm{even}}$ vanishes for $\theta=0$ and $\pi$ because it is
proportional to $m_xm_z$, whereas $\delta S_z^{\mathrm{even}}$ and
$\delta S_y^{\mathrm{odd}}$ are generally finite for
$\mathbf{m}\parallel\pm x$.

For direct comparison with the first-principles angular dependence,
Eq.~(\ref{eq:xz_angular_poly}) can be recast into an exactly equivalent
compact form using $m_x^2+m_z^2=1$ and standard trigonometric identities:
\begin{equation}
\begin{aligned}
\frac{\delta S_x^{\mathrm{even}}}{E_x}
&=
A_2\sin2\theta
+
A_4\sin4\theta
+
A_6\sin6\theta,
\\
\frac{\delta S_z^{\mathrm{even}}}{E_x}
&=
B_2\cos^2\theta
+
B_4\cos^4\theta
+
B_6\cos^6\theta,
\\
\frac{\delta S_y^{\mathrm{odd}}}{E_x}
&=
C_1\cos\theta
+
C_3\cos^3\theta
+
C_5\cos^5\theta.
\end{aligned}
\label{eq:xz_fit}
\end{equation}
Here $A_{2,4,6}$, $B_{2,4,6}$, and $C_{1,3,5}$ are the coefficients used to
fit the first-principles results. They are related to the coefficients of the
symmetry-constrained polynomial expansion by
\[
\begin{aligned}
A_2
&=
\frac{1}{2}\alpha_2
+
\frac{1}{4}\alpha_4
+
\frac{5}{32}\alpha_6,
&
\\A_4
&=
-\frac{1}{8}\alpha_4
-\frac{1}{8}\alpha_6,
&
A_6
&=
\frac{1}{32}\alpha_6,
\\
B_2
&=
\beta_2+\beta_4+\beta_6,
&
\\B_4
&=
-\beta_4-2\beta_6,
&
B_6
&=
\beta_6,
\\
C_1
&=
\gamma_1+\gamma_3+\gamma_5,
&
\\C_3
&=
-\gamma_3-2\gamma_5,
&
C_5
&=
\gamma_5.
\end{aligned}
\]
Equation~(\ref{eq:xz_fit}) therefore introduces no additional approximation,
but constitutes an exactly equivalent parametrization of the symmetry-allowed
expansion along the $xz$ rotation path. Its leading terms reduce to
$\delta S_x^{\mathrm{even}}\propto\sin2\theta$,
$\delta S_z^{\mathrm{even}}\propto\cos^2\theta$, and
$\delta S_y^{\mathrm{odd}}\propto\cos\theta$, respectively, whereas the
higher-order terms account for the pronounced deviations from the lowest-order
angular dependence observed in the first-principles results.

\section{Effect of broadening}
\label{app:broadening}

The Kubo response depends quantitatively on the phenomenological broadening parameter $\Gamma$, which accounts for the finite lifetime of the electronic states within the constant-broadening approximation. In the main calculations, we adopt $\Gamma=10$~meV. Using the convention $\tau=\hbar/(2\Gamma)$, this value corresponds to a characteristic relaxation time of approximately $33$~fs. Femtosecond carrier-relaxation times are commonly encountered in transport studies of two-dimensional systems \cite{Abdukayumov2025SpinCharge,Kempt2024PtSe2}. More directly relevant to the present materials, first-principles calculations for monolayer \ce{Fe3GeTe2} reported phonon-mediated scattering rates of approximately $62$--$117$~ps$^{-1}$ at 300~K, corresponding to carrier lifetimes on the order of $10$~fs~\cite{Badrtdinov2023FGTTransport}. Furthermore, $\Gamma=10$~meV has been employed in previous first-principles studies of the Edelstein effect in two-dimensional systems \cite{Li2020PNAS,Pan2026PtBi2,Pan2026VTe2}. These considerations indicate that $10$~meV provides a reasonable reference value for the present comparative calculations. We nevertheless regard $\Gamma$ as a phenomenological parameter rather than a material- and energy-resolved scattering rate.

To explicitly examine its influence, we calculate the symmetry-allowed Edelstein responses of monolayer \ce{Fe3GaTe2} and \ce{Fe3GeTe2} as a function of $\Gamma$. The electric field and magnetization are both fixed along the $+x$ direction, for which the allowed components are $\delta S_z^{\mathrm{even}}$ and $\delta S_y^{\mathrm{odd}}$. The energy is fixed at $E-E_F=-0.40$~eV, where sizable responses are obtained, and the results are shown in Fig.~\ref{fig9}. In both compounds, $\delta S_z^{\mathrm{even}}$ increases rapidly as $\Gamma$ is reduced, whereas $\delta S_y^{\mathrm{odd}}$ displays a much weaker dependence and does not show the same divergent behavior.

This contrasting behavior can be understood directly from the Kubo
expressions~\cite{Freimuth2014,Pan2026VTe2}. In the small-$\Gamma$ limit,
the time-reversal-even contribution reduces to
\begin{equation}
\chi_{ij}^{\mathrm{even}}
\simeq
-\frac{e\hbar}{2\Gamma N}
\sum_{\mathbf{k},n}
\delta(\varepsilon_{n\mathbf{k}}-E_F)
\langle n\mathbf{k}|\hat{S}_i|n\mathbf{k}\rangle
\langle n\mathbf{k}|\hat{v}_j|n\mathbf{k}\rangle ,
\label{eq:broad_even}
\end{equation}
which contains an explicit $1/\Gamma$ factor. The time-reversal-even response is therefore dominated by the intraband contribution at the Fermi surface and grows with increasing carrier lifetime. This behavior is consistent with the strong enhancement of $\delta S_z^{\mathrm{even}}$ toward small $\Gamma$ in Fig.~\ref{fig9}.

In contrast, the time-reversal-odd contribution approaches, in the same
limit, the intrinsic interband form
\begin{equation}
\begin{aligned}
\chi_{ij}^{\mathrm{odd}}
\simeq{}&
-\frac{2e\hbar}{N}
\sum_{\mathbf{k}}
\sum_{n\in\mathrm{occ.}}
\sum_{\substack{m\in\mathrm{unocc.}\\m\neq n}}
\\
&
\frac{
\mathrm{Im}\!\left[
\langle n\mathbf{k}|\hat{S}_i|m\mathbf{k}\rangle
\langle m\mathbf{k}|\hat{v}_j|n\mathbf{k}\rangle
\right]
}{
(\varepsilon_{n\mathbf{k}}-\varepsilon_{m\mathbf{k}})^2
}.
\end{aligned}
\label{eq:broad_odd}
\end{equation}
which contains no explicit dependence on $\Gamma$. Accordingly, the time-reversal-odd response does not exhibit the $1/\Gamma$ enhancement of the even contribution. The residual variation visible at finite $\Gamma$ arises because the calculations are not necessarily in the asymptotic small-$\Gamma$ limit over the entire range considered, particularly when the broadening becomes comparable to relevant interband energy separations. This behavior also reflects the different microscopic origins of the two responses: the even component is primarily intraband and Fermi-surface controlled, whereas the odd component originates from interband coherence between nearby electronic states.

Although the absolute magnitude of the calculated spin accumulation depends on $\Gamma$, the symmetry content of the response is unaffected. In particular, varying $\Gamma$ does not alter the allowed tensor components, their transformation properties under magnetic symmetry operations, or the angular parity relations derived in Appendix~\ref{app:angular}. The broadening analysis therefore supports the use of $\Gamma=10$~meV as a common reference for comparing \ce{Fe3GaTe2} and \ce{Fe3GeTe2}, while the main symmetry-based and electronic-structure trends discussed in this work remain unchanged.

\bibliography{FGT}

\end{document}